%% file: text.tex
\documentclass[twocolumn]{aastex631}

\newcommand{\pow}[1]{\times 10^{#1}}

\usepackage{array}
\usepackage{textcomp}
\usepackage{amsmath}
\usepackage{booktabs}
\usepackage{makecell}
\usepackage{tabularx}
\usepackage{multirow}
\usepackage{tablefootnote}

\begin{document}



\title{BICEP/\emph{Keck} XIX: Extremely Thin Composite Polymer Vacuum Windows for BICEP and Other High Throughput Millimeter Wave Telescopes}
\input{authors.tex}



\begin{abstract}
    Millimeter-wave refracting telescopes targeting the degree-scale structure of the cosmic microwave background (CMB) have recently grown to diffraction-limited apertures of over 0.5 meters. These instruments are entirely housed in vacuum cryostats to support their sub-kelvin bolometric detectors and to minimize radiative loading from thermal emission due to absorption loss in their transmissive optical elements. The large vacuum window is the only optical element in the system at ambient temperature, and therefore minimizing loss in the window is crucial for maximizing detector sensitivity. This motivates the use of low-loss polymer materials and a window as thin as practicable. However, the window must simultaneously meet the requirement to keep sufficient vacuum, and therefore must limit gas permeation and remain mechanically robust against catastrophic failure under pressure. We report on the development of extremely thin composite polyethylene window technology that meets these goals. Two windows have been deployed for two full observing seasons on the BICEP3 and BA150 CMB telescopes at the South Pole. On BICEP3, the window has demonstrated a 6\% improvement in detector sensitivity.
\end{abstract}

\section{Introduction}

The cosmic microwave background (CMB) is the afterglow from the Big Bang, and as such, is an excellent tool to  explore cosmological models \citep{Penzias_Wilson_1965,Planck_Cosm_2020}. CMB photons---emitted as the opaque primordial plasma cooled into transparent neutral gas---carry information about the conditions of the early universe. The CMB has a near-perfect black body spectrum, ruling out a steady state expanding universe, and the temperature deviations on that spectrum provide strong evidence for the standard model of cosmology, $\Lambda$CDM \citep{FIRAS1994,WMAP2003,Planck_Cosm_2020}. There remain unresolved questions about the early universe (such as the horizon and flatness problems) that paradigms like inflation attempt to explain. Currently, several experiments are aiming to constrain the parameters of inflation by mapping the CMB. A very small excess of primordial ``B modes" (curl in the polarization field of the CMB) at degree scales is an expected prediction from Inflation in the early universe \citep{Kamionkowski_Kovetz_2016}.

As constraints on cosmological parameters advance, there is a need to improve sensitivity of the telescopes making these measurements. There are broadly two ways to accomplish this: increase detector counts while keeping the per-detector sensitivity the same or decrease the optical loading thus increasing the per-detector sensitivity. These two goals lead to conflicting physical designs. Increasing the detector count requires a larger aperture size, which leads to thicker optical components. Since the optical emission from a component increases with its thickness, larger apertures tend to degrade the per-detector sensitivity.

The BICEP/\emph{Keck} series of experiments are small-aperture on-axis refracting telescopes, designed to optimize high optical throughput at degree scales. This strategy has been successful in placing leading constraints on the millimeter polarized signal at degree scales for the past twenty years \citep{BK18}. Each generation of BICEP/\emph{Keck} receivers has increased the number of detectors by roughly an order of magnitude: BICEP1 had a 250 mm aperture for 49 detector pairs at 95, 150, and 220 GHz \citep{BICEP1b}; BICEP2 (150 GHz) and the \emph{Keck} Array of five receivers (150 GHz, later adding 95, 220 and 270 GHz) had 256 detector pairs each for their 264 mm apertures \citep{BKIII,Kernasovskiy2012_Keck,Staniszewski2012_Keck}; the currently operating BICEP3 (95 GHz) has 1200 detector pairs in a 520 mm aperture \citep{bicep3}. A single BICEP Array receiver---the set of receivers currently being deployed which range in frequency from 30 to 300 GHz---has between 100 and 4000 detector pairs (depending on the designed frequency) for a 560 mm aperture \citep{Moncelsi2020_BA,Schillaci2023_BA150,Nakato2024}. 

To accommodate the order of magnitude increase in detector count between the recent stages of BICEP receivers the aperture size of the receivers has doubled from 264 mm to between 520--560 mm \citep{Keck2015B,Moncelsi2020_BA}. The readout electronics and transition edge detectors are carefully designed so that the receivers are photon-noise limited. Therefore reducing in-band optical load can have a profound impact on a given BICEP receiver's per-detector sensitivity.
\begin{figure*}[t]
    \centering
    \includegraphics[width=0.47\textwidth]{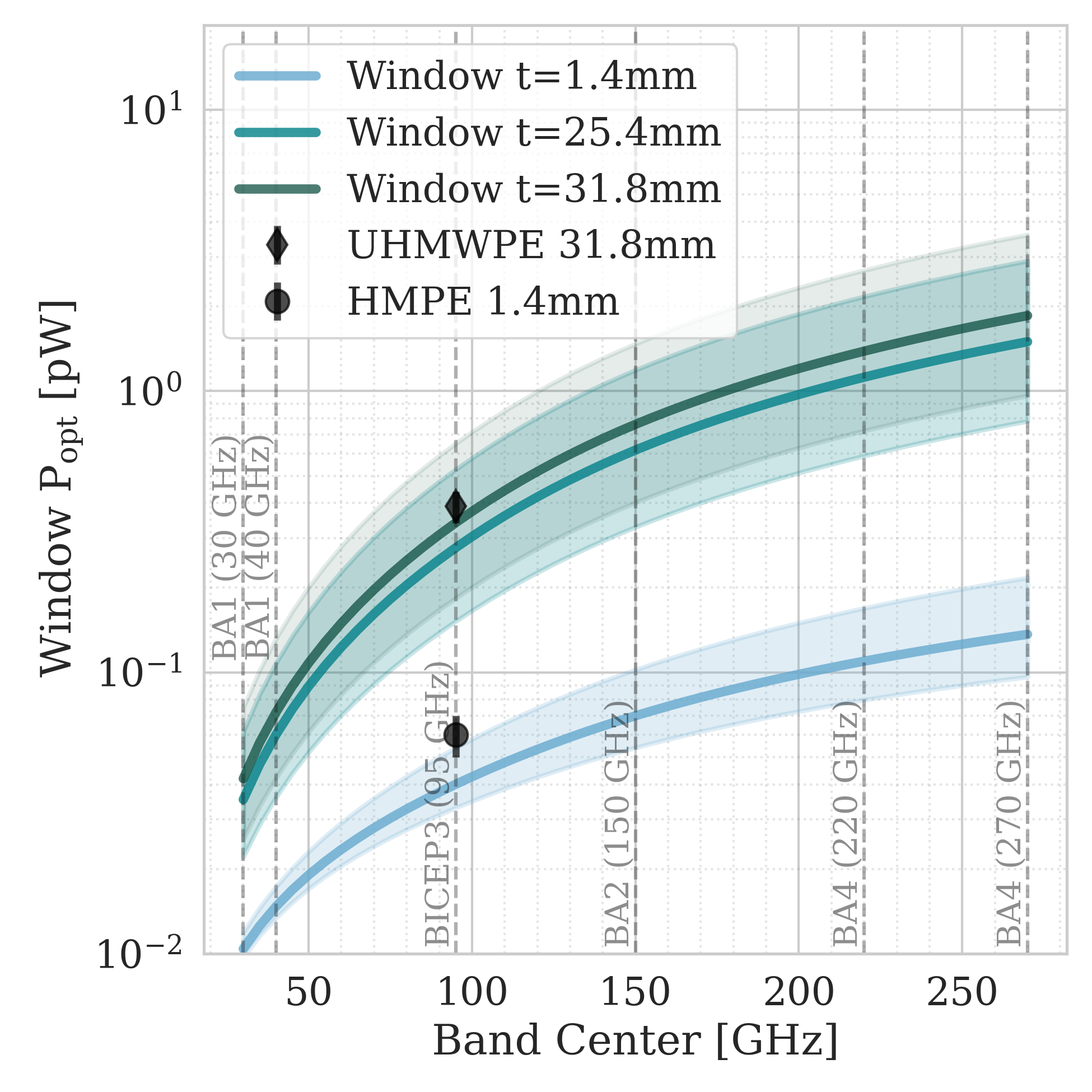}
    \includegraphics[width=0.47\textwidth]{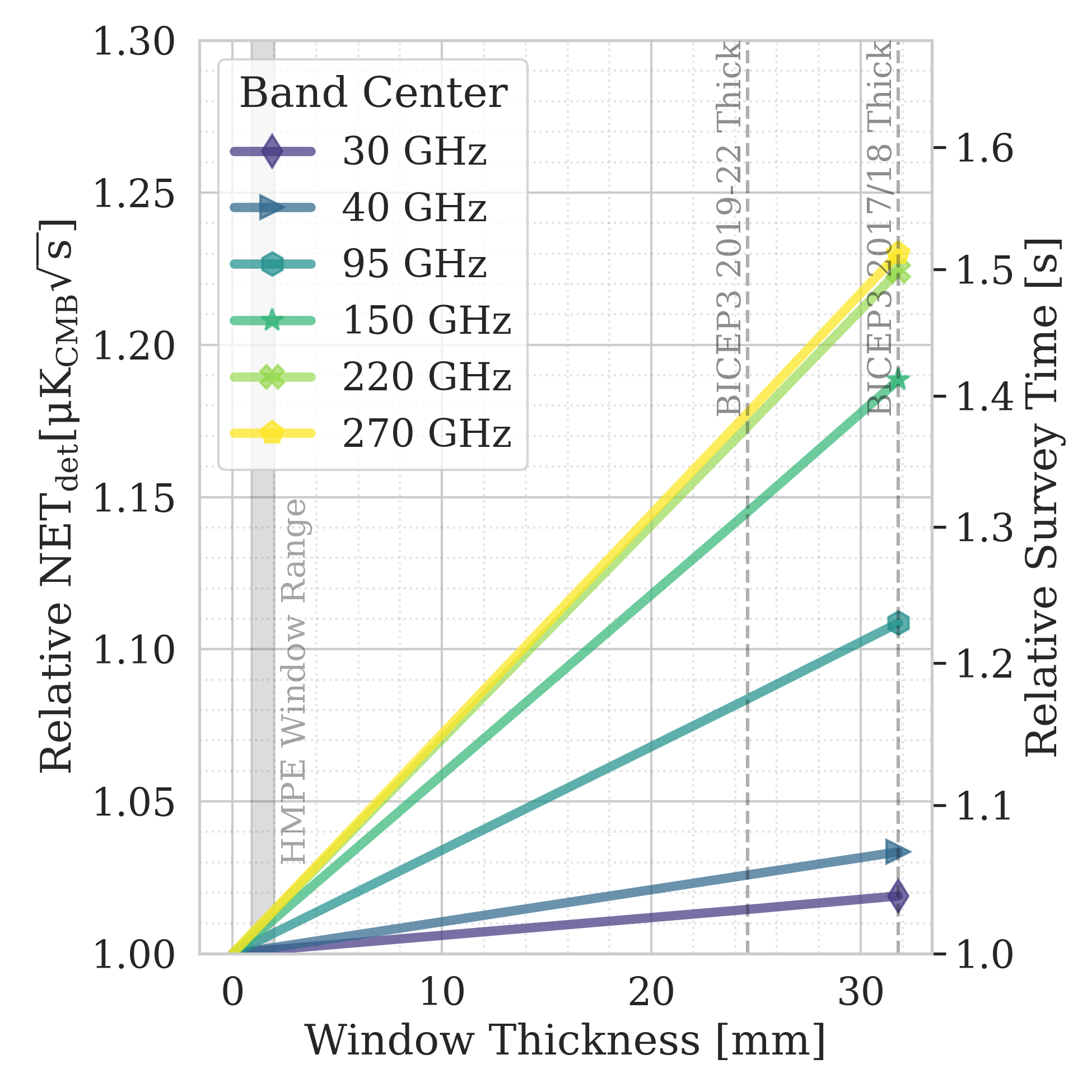}
    \caption{[Left] Modeled optical loading from a window with thicknesses of 1.4 mm, 25.4 mm, and 31.8 mm, at band centers from 30 to 270 GHz (fractional bandwidth of 0.25) through a BICEP-style instrument. Black points are measured optical loading on BICEP3 from a 31.8 mm (1.25") UHMWPE spare window and the 1.4 mm laminate window deployed on BICEP3 for the 2023 season. Shaded bands are uncertainty on the loss properties of the window, with \(\tan\delta\) between 0.6$\pow{-4}$ to 2.4$\pow{-4}$. [Right] Modeled relative noise equivalent temperature per detector (NET$_{\text{det}}$) and relative survey time increase compared to an instrument without a window, for each of the BICEP Array observing band center frequencies.} 
    \label{fig:intro_load_net}
\end{figure*}
The in-band radiative load emitted by the optical elements is the product of their absorptive loss and physical temperature. 
The absorption loss of a material is directly related to the complex permittivity ($\varepsilon = \varepsilon' - i \varepsilon''$) of a material. Absorption is discussed numerically throughout this paper as a material's loss tangent, which is the ratio of the real part over the complex part of the permittivity, denoted as $\tan\delta$. The permittivity of a material varies with temperature, and the $\tan\delta$ at millimeter wavelengths of many polymers decreases with cryogenic cooling, reducing the absorption and emission of a polymer optic in-band \citep{Frank1977,Schnabel2014}. However, the reduction in optical emission at lower temperature is driven primarily by the lower physical temperature rather than the lower loss \citep{Lamb1996}.

The entire instrument is placed within a vacuum cryostat to cool both the optics and the sub-kelvin detectors. However, the cryostat requires a transmissive window to observe external signals, which must remain in contact with the ambient environment. Vacuum windows must have low transmission loss in band and be strong enough to withstand the force of atmospheric pressure. Previously, millimeter-wave vacuum windows have been made of ceramics [such as fused-silica \citep{piper2021}], plastic foams [including various formulations of Zotefoam \citep{Keck2015B}], or bulk plastic materials [primarily high density polyethylene (HDPE) or ultra high molecular weight polyethylene (UHMWPE) \citep{bicep3,delessandro2018}]. A more detailed history of the development of millimeter vacuum windows can be found in \cite{Barkats2018}.

The larger aperture sizes of modern CMB receivers, however, limit the materials available for use. BICEP3 and BICEP Array have nominal clear apertures at the window of 730 mm, while the full outer diameter of the window is 900 mm. Plans for the CMB-S4 Small Aperture Telescopes (SATs) and Large Aperture Telescope (LATs) included similarly sized windows \citep{abazajian2019cmbs4}. Ceramics become prohibitively expensive at those scales. Plastic foams would become impracticably thick and lossy: the primary advantages of the foam windows are their low index of refraction and low transmission loss, but they have the disadvantage of being very weak. A suitably thick foam window would include many laminated layers, which would make their loss properties significantly worse. For example, the 150 GHz \emph{Keck} receiver windows were 120 mm thick and estimated to cause $\lesssim$2\% transmission loss \citep{Keck2015B}; the doubled aperture size for new BICEP receivers would require a foam window four times thicker. A window with such high transmission loss ($\lesssim$8\%) would add approximately 3 pW optical load to a 150 GHz detector, swamping out all other sources of optical power at that frequency.

Bulk plastics, being relatively weak though still stronger than plastic foams, must also become significantly thicker to maintain a suitable safety factor for a vacuum vessel (discussed further in Section \ref{sec:strength}). For BICEP3, a 95 GHz receiver, we previously estimated that over half of the instrument loading originated from the 31.8 mm thick HDPE slab window \citep{bicep3}. That window has since been replaced by the thin window presented in this paper.

In Figure \ref{fig:intro_load_net} we model the optical loading on a large aperture BICEP-style receiver for six observed bands and different window thicknesses. We also show the relative change in per-detector white noise and survey time for different window thicknesses. For the lowest frequency band, at 30--40 GHz, decreasing the window thickness only yields marginal improvements in white noise and mapping speed. For bands above 40 GHz, however, a substantial reduction in window thickness from the nominal could potentially decrease noise by tens of percent. At the high frequencies in the 270 GHz band, such a window thickness reduction could potentially decrease the survey time by as much as 50\%. The window is the only element in the optical chain where improvement would produce such a significant return, primarily due to the window's relatively high temperature. 

\begin{figure*}[t]
    \centering
    \includegraphics[width=0.9\linewidth]{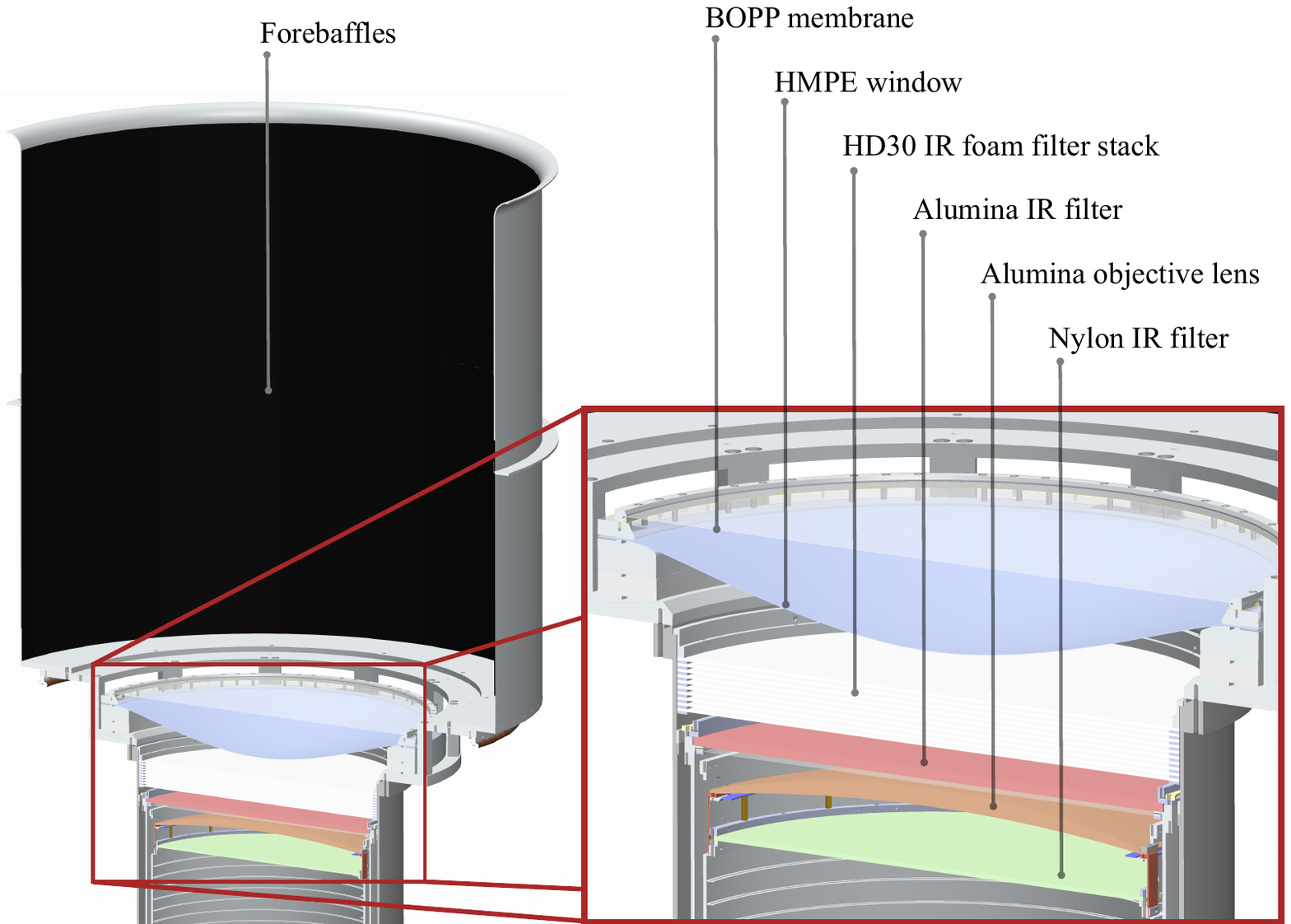}
    \caption{A cutaway of the top portion of the BICEP3 cryostat. Displayed from top to bottom are the forebaffles, BOPP membrane (to contain warm circulated air above the window), thin HMPE window, HD30 IR foam filter stack, alumina IR filter, alumina objective lens, and first nylon IR filter. The rest of the cryostat is described in Figure 6 of \cite{bicep3}. The thin HMPE window in this model is deflected 75 mm, with the foam filters remaining 45 mm away from the window.}
    \label{fig:bicep3_cutaway}
\end{figure*}

We have achieved an order of magnitude reduction in window thickness using a special form of polyethylene we call high modulus polyethylene (HMPE, commercial name Dyneema \citep{Dyneema} or Spectra) laminated with low density polyethylene (LDPE) to generated a polyethylene composite material.

As of 2025, we have deployed three thin laminate windows on BICEP/\emph{Keck} receivers; one as a retrofit on the 95 GHz BICEP3, one on the new 150 GHz BICEP Array receiver (hereafter BA150) and another on the new 220/270 GHz BICEP Array receiver (hereafter BA220/270). Section \ref{sec:design} of this paper discusses the design considerations and mechanical tests for these windows. Section \ref{sec:optical_tests} explores in-lab validation of optical properties such as polarization and anti-reflection. Section \ref{sec:BICEP3_char} reviews optical characterizations performed on the deployed window before and after installation on BICEP3. We then conclude with final thoughts and next steps.

\section{Design and Mechanical Tests} \label{sec:design}The high strength of the HMPE allows for window thicknesses of around a millimeter---similar to the wavelength of light we are observing, which  will become relevant for the optical properties of the window. The LDPE melts at lower temperature than the HMPE, allowing for the laminating material to fully surround the HMPE fibers without compromising the strength carrier's integrity \citep{Vasile2005,Eiben2022}. Because the composite is entirely polyethylene, one should expect the window to appear homogeneous to millimeter light, and therefore have similar optical properties---such as index of refraction, loss and scattering---as standard bulk polyethylene. 

There are three parameters that influence the in-band optical emission of a material: the physical temperature, the absorption coefficient, and its thickness. The temperature is not a parameter than can be changed as the window is in thermal equilibrium between the ambient temperature outside and the cold optics inside the cryostat. Similarly, the absorption coefficient cannot be changed; it is a property of the material itself (see \cite{Barkats2018} for further discussion). Therefore the thickness is the only parameter available to reduce the optical loading from the window.
Unfortunately, bulk plastic windows tend to be very thick due to bulk polyethylene's low strength. 
Therefore, to decrease the thickness of the window, we must increase the strength of the materials we use. 

HMPE fibers have an ultimate tensile strength over 150 times stronger than bulk HDPE or bulk UHMWPE \citep{Barkats2018, Eiben2022}. However, the HMPE fibers are woven into a fabric, which is not naturally vacuum tight. We must add an additional processing step, laminating the HMPE with another material that should be optically matched to the strength carrier. As LDPE is also a polyethylene, the millimeter optical properties are expected to be very similar. We add LDPE layers between HMPE layers and then the LDPE, having a lower melting point than HMPE, can be melted to laminate the fabric layers and fill the voids between fibers \citep{Vasile2005}. This makes the composite of HMPE and LDPE vacuum tight \citep{Barkats2018, Eiben2022}.



As shown in Figure \ref{fig:bicep3_cutaway}, the window is the first optical element facing the sky. Above the window are ``forebaffles" which terminate sidelobe and scattered pickup outside the nominal 27.4$^{\circ}$ field of view of the telescope at a stable, warm temperature \citep{bicep3}. Below the window are a stack of physically separated thin polyethylene foam sheets (Zotefoam HD30) which act as an IR filter \citep{Choi_2013, Goldfinger2022}. These thin laminate windows must fulfill a variety of criteria to avoid negatively impacting our sensitive instruments. Mechanically, the window should not allow excessive amounts of gas into the cryostat (Section \ref{sec:perm}), must withstand the force of atmospheric pressure with a reasonable safety factor (Sections \ref{sec:strength} and \ref{sec:hydro_stat}), and must not deflect too far inwards to avoid contacting the IR foam filter below them (Section \ref{sec:creep}).

\subsection{Manufacturing} \label{sec:manufact}
The thin windows are composite polyethylene laminates. The HMPE comes in a woven fabric on rolls 1500 mm wide, out of which 900 mm diameter circles are cut with a hot wire; 900 mm is the outer diameter of the frames which hold these windows on the cryostat. At least three layers of HMPE are included in every window, and each layer is rotated 30 degrees from the one below. The number of HMPE layers is primarily set by the  maximum allowed deflection and accounts for the expected creep it will experience over its lifetime (more details in Sections \ref{sec:strength} and \ref{sec:creep}). LDPE layers surround the HMPE: we control the final laminate thickness with the number and initial thickness of LDPE layers. At least four layers of 100 $\mu$m (4 mil) LDPE are required to fully laminate the three layers of HMPE. 

\begin{figure}[ht]
    \centering
    \includegraphics[width=\linewidth]{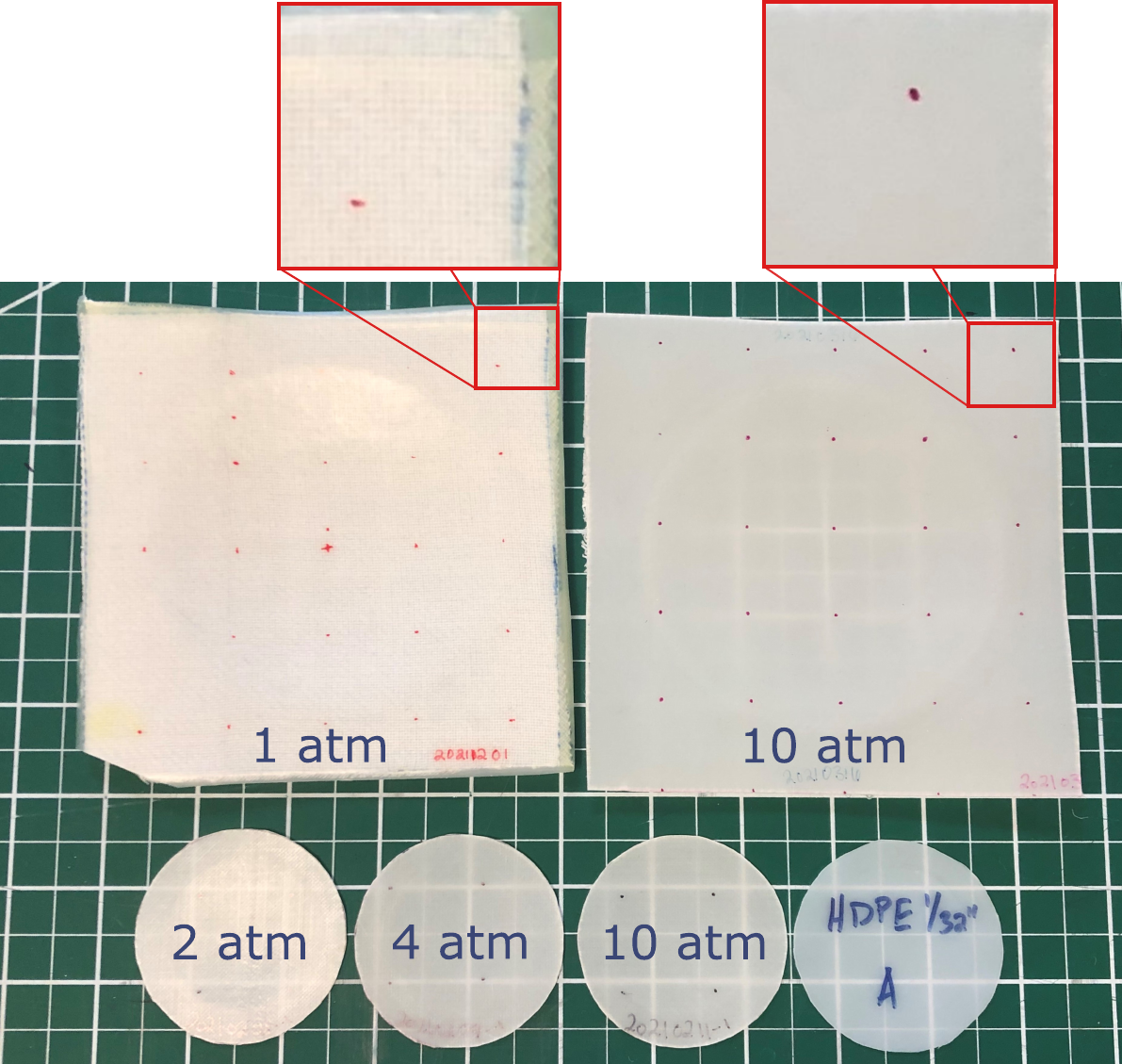}
    \caption{Photo of several small scale laminated window samples, labeled with the pressure at which they were laminated in atmospheres (atm). The top two samples were used in the small-scale permeation tests in Figure \ref{fig:perm_rates}; the laminate recipes are identical apart from the difference in lamination pressure. The weave of the HMPE fibers remains visible in the  1 atm laminate, whereas the 10 atm laminate appears more homogeneous, like the HDPE sample below. he row of three smaller HMPE laminate samples at different pressures (homogeneity of the laminates requires pressures at or above 4 atm) and a solid extruded HDPE sheet have the same thickness. Note that the grid pattern on the mat below the samples laminated at lower pressure is blurred by light scattering associated with the incomplete filling of the HMPE weave with LDPE.}
    \label{fig:lam_qual}
\end{figure}

Lamination is processed at high pressure (10 atm) to fully surround the HMPE fibers, which is visually demonstrated in Figure \ref{fig:lam_qual}; Sections \ref{sec:perm} and \ref{sec:antiref} both describe  mechanical and optical difficulties encountered with poor lamination. We determined that lamination above 4 atm was necessary for homogeneity, and that higher pressure lamination was more likely to keep the appropriate laminating force during a cycle. The highest pressure we can achieve is 10 atm, so all laminates were baked at 10 atm unless otherwise indicated. The lamination temperature must remain between 130$^\circ$C and 140$^\circ$C to melt the LDPE around the HMPE without affecting the HMPE's strength: we typically aim for 135$^\circ$C to 137$^\circ$C. When appropriate we distinguish between windows by their manufacturing processes, i.e., the pressure and/or temperature of the lamination. The design of the optics autoclave we use for this heat compression process is described further in \cite{Eiben2022}.

The minimum thickness of any window is set by the strength of the laminate, which is discussed in Section \ref{sec:strength}. The actual designed thickness is set by the desired anti-reflection properties, discussed further in Section \ref{sec:antiref}.

\subsection{Strength} \label{sec:strength}
\begin{figure}[b]
    \centering
    \includegraphics[width=\linewidth]{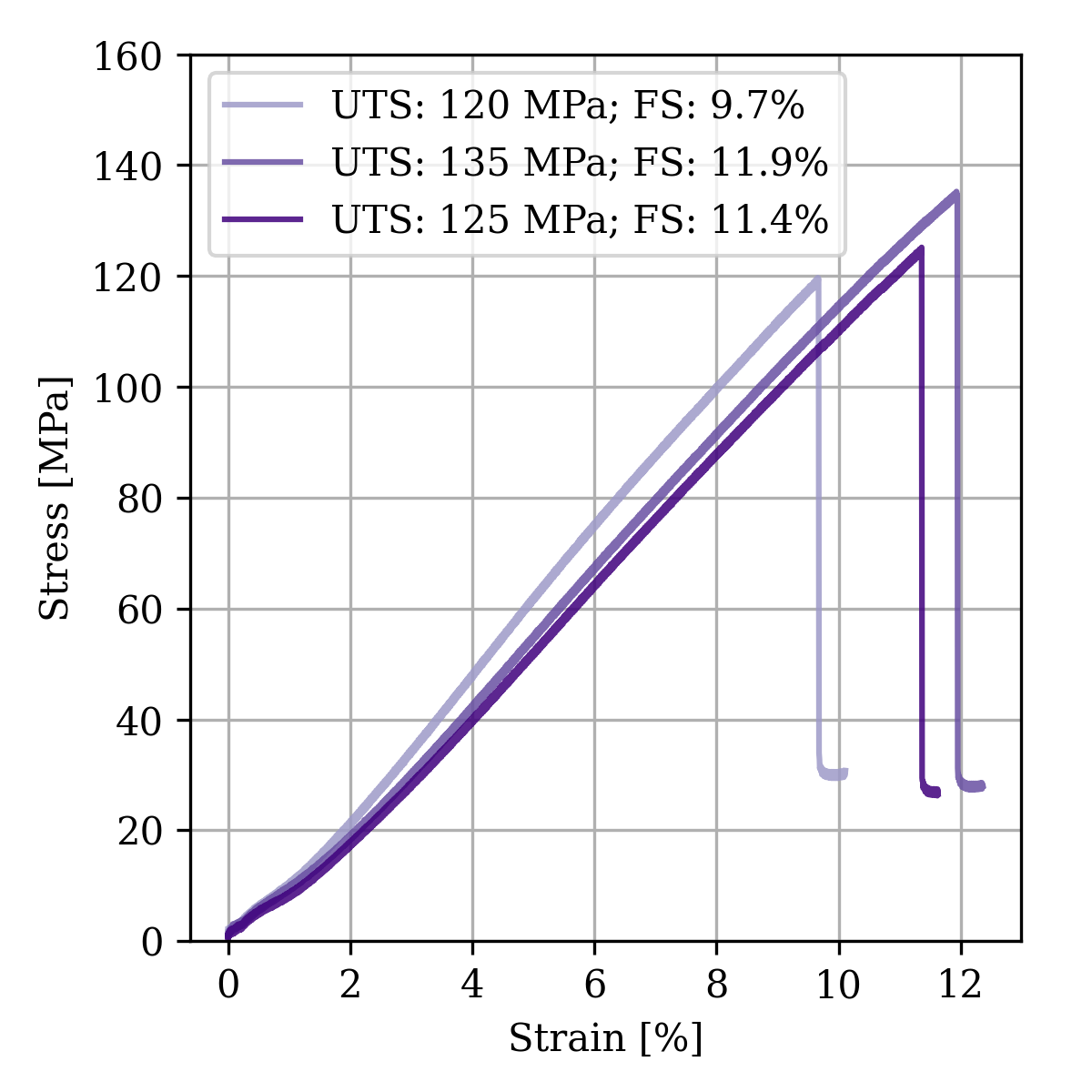}
    \caption{Tensile strength tests of three samples from a high pressure (10 atm, 135$^{\circ}$C) HMPE laminate. The stress is calculated by dividing the measured force by the initial cross sectional area of the sample, while the strain is the percent change in elongation. The ultimate tensile strength (UTS) and the failure strain (FS) for each sample are reported in the legend, and summarized and compared to other polyethylenes in Table \ref{tab:strength}.}
    \label{fig:tensile_tests}
\end{figure}

\begin{table*}[t!]
\caption{Strength Properties of Bulk HDPE, UHMWPE, and HMPE Laminates}
\begin{tabularx}{\textwidth}{*{5}{>{\centering\arraybackslash}X}}
    \hline \hline
         Window Material & Ultimate Tensile Strength & Elastic Modulus & Strain at break & Thickness for Safety Factor of 3 \tablenotemark{a}\\
        & (MPa) & (GPa) & (\%) & (mm)\\
        \hline
        HDPE & 20 -- 25 & 0.9 & 10 & 17.7\\
        UHMWPE & 22 -- 40 & 0.7 & 30 & 13.6\\
        HMPE Laminate & \multirow{2}{*}{120 -- 135} & \multirow{2}{*}{1.1 -- 1.3} & \multirow{2}{*}{9 -- 12} & \multirow{2}{*}{1.4} \\
        (10 atm) &  &  &  &  \\
        \hline
\end{tabularx}
\tablenotetext{a}{Estimated using Equation \ref{eq:max_stress} and the lower of the reported ultimate tensile strengths.}
\tablerefs{\cite{Barkats2018,delessandro2018}}
    \label{tab:strength}
\end{table*}
In order to dimension these new thin windows correctly, we first start with the measured mechanical properties: the elastic modulus, the ultimate tensile strength, and the strain at break. We measure these mechanical properties using a tensile strength machine that collects force versus elongation data over the range of extension of the material up to failure. Those data are then converted into engineering stress (force over the initial cross sectional area) and strain (percent elongation). Example tensile strength measurements for a 10 atm 135$^{\circ}$C HMPE laminate are shown in Figure \ref{fig:tensile_tests}. These measurements are also presented in Table \ref{tab:strength} for the HMPE laminate material, along with literature values for the bulk HDPE and UHMWPE found in \cite{Barkats2018,delessandro2018}.

The ultimate tensile strength is the maximum stress a material can withstand before failure, which may vary slightly between samples due to measurement uncertainty, particularly uncertainty in the cross sectional area. The failure strain is the percent elongation a material can withstand before failure. The slope of the stress/strain curve is the elastic modulus, which is particularly useful to predict how a material will deform (strain) in response to a load (stress).

With these properties we can compute the expected plastic deformation and stress of a window under atmospheric load. Assuming we are in the large angle deflection regime ($\delta$) where the deflection is large compared to the thickness of the window, the initial elastic deformation is described by the flat plate equation:
\begin{equation}
    \delta = K \Big(\frac{\Delta P \, R^4}{E \,h}\Big)^{(1/3)} .\label{eq:delta} 
\end{equation}

While the maximum stress experienced under such deflection is given by:
\begin{equation}\label{eq:max_stress}
		\sigma_{max} =  Z \left(\frac{ E \Delta P^2 R^2}{h^2}\right)^{1/3}
\end{equation}
where $K$ or $Z$ are a constant depending on the location of interest of the plate ($K$ = 0.662, $Z = 0.423$ for the center), 
$R$ is the radius of the plate, $E$ is the flexural modulus (assumed equal to the elastic modulus in most material), $\Delta P$ is the pressure difference (equal to one atmosphere), and $h$ is the thickness of the material, all in SI units \citep{TimoshenkoStephen1959Topa}.

The ratio of the window’s ultimate tensile strength to the predicted stress is the safety factor. We can iterate on the window thickness to achieve a desired safety factor, which we set to be at least three. Section \ref{sec:hydro_stat} presents a detailed motivation for this choice. One way to compare the polyethylene window materials is to calculate the thickness required for each polyethylene window material to reach the required safety factor, using the reported material properties in Table \ref{tab:strength} and Equation \ref{eq:max_stress}. We use the lower value in the range of reported ultimate tensile strengths for a given material to determine the minimum safe thickness reported in Table \ref{tab:strength}. Both bulk polyethylenes have similar lower limits on the ultimate tensile strength and therefore require an order of magnitude thicker material to achieve the same safety factor as the HMPE window.



It is important to note that these deflection, stress, and strain calculations use the full thickness of the laminate even though a significant fraction of the thickness is taken up by the much weaker LDPE. As such, the safety factor thicknesses reported in Table \ref{tab:strength} are a lower limit. Previously, we reported the strength of laminates using the HMPE fiber cross section \citep{Barkats2018}. We find that recent laminates have the same high ultimate tensile strength (at or above 3 GPa) when normalizing tensile results to the fiber cross section, which indicates that the lamination process does not reduce the HMPE's strength.
To confirm that the thin laminate windows have an appropriately high margin to their ultimate strength, we conducted an empirical test of the laminate's strength, under the clamping boundary conditions that a deployed window would experience, which we discuss in the next section.

\subsection{Safety Factor Measurement} \label{sec:hydro_stat}

Traditional tensile tests like those shown in Figure \ref{fig:tensile_tests} will only stress the woven fabric in a laminate in a single direction (the direction of the pulling force). However, a three layer laminate will have HMPE fibers running in six separate directions over a circular aperture (each layer is rotated 30 degrees from the one below it). A tensile test does not accurately capture the stress conditions of an actual window. The scaling of tensile tests to the full scale conditions thus remains uncertain: the maximum stress and safety factor calculations presented in Section \ref{sec:strength} rely on the thin plate analytical relationship, which assume homogeneous material, and may not fully apply to composites. To mitigate this concern, we performed a full scale window failure measurement.

For safety, we performed this test hydrostatically. As water is nearly incompressible, minimal energy will be stored within the fluid as pressure is increased; energy is only stored within the deflected window and surrounding enclosure. Given the safety factors of 3 calculated in Table \ref{tab:strength}, we used standard utility water pressure (roughly 60 psi or 4 atm)  as that was expected to exceed the ultimate tensile strength of these windows and produce a failure.

\begin{figure}[t]
    \centering
    \includegraphics[width=\linewidth]{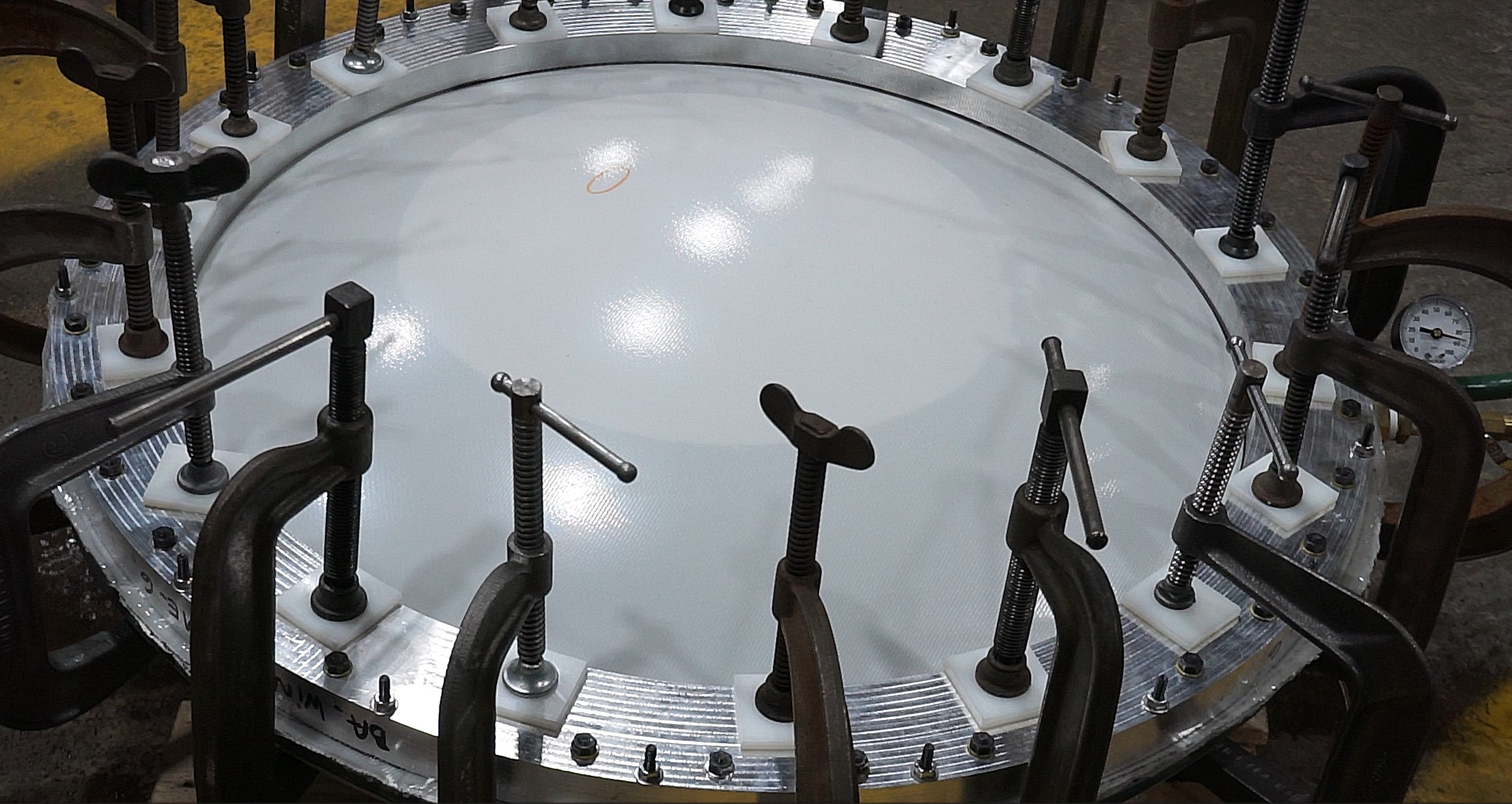}
    \caption{Photo of the window at the highest pressure achieved during the hydrostatic test. The pressure gauge (center right) reads approximately 85 psi (5.7 atm, or 586 kPa), and there is water leaking out around the gasket seal on the left hand side.}
    \label{fig:hydrostat_high_p}
\end{figure}

The BICEP Array frame and clamp were designed specifically to hold thin windows with a knurled pattern on the outer radius, with bolts running through the knurling to maximize concentration of force over relatively small points \citep{Barkats2018}. This knurled pattern on the frame and clamp works well but results in an unique window clamping configuration. We therefore used an actual frame and clamp for the hydrostatic test. Preexisting flange holes for circulating gas between the window and BOPP membrane were used to push the water into the chamber above the window. We sealed the water chamber with a 1/2" aluminum blanking plate and two 3/16" thick blended EPDM gaskets. The chamber was clamped with two sets of correctly torqued 1/4" bolts and C-clamps for additional clamping force, which are visible in a ring around the window in Figure \ref{fig:hydrostat_high_p}.

\begin{figure*}[ht!]
    \centering\includegraphics[width=0.8\textwidth]{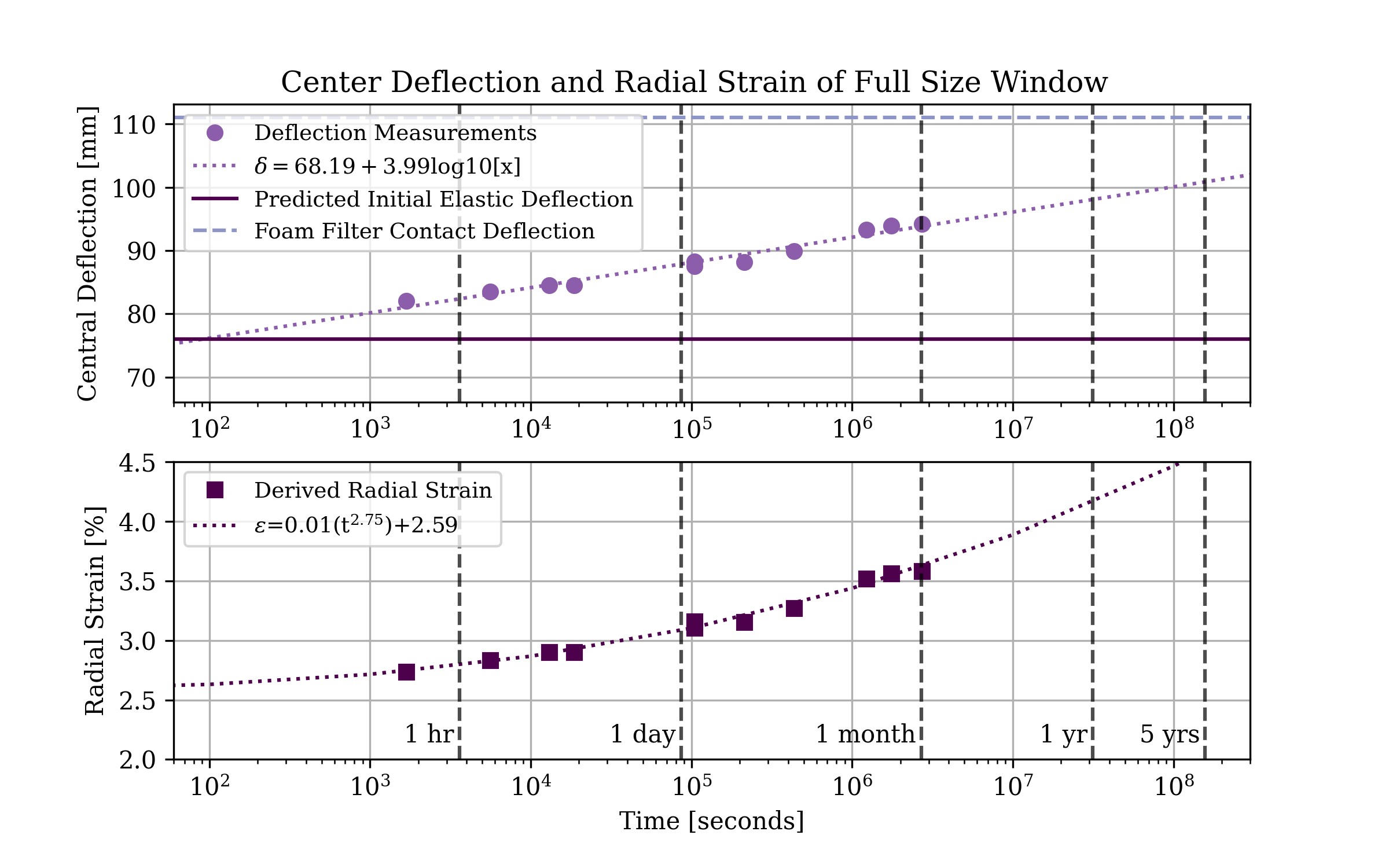}
    \caption{[Top] Measured center deflections of a window on the a BICEP Array scale cryostat. The purple solid line denotes the predicted elastic deflection from the plate equation Eq. \ref{eq:delta} using a radius of 400 mm. The light purple dashed line denotes the deflection at which contact would be made with the foam infrared filters behind the window. [Bottom] Estimated radial strain using Eq. \ref{eq:rstrain1} and the measured deflections.}
    \label{fig:creep_def}
\end{figure*}
We reached 68 psi (469 kPa) in the hydrostatic test using only the standard utility water supply. At that point we took a side-on picture of the deflected window, and estimated the deflection (and therefore radial strain) of the window (see Section \ref{sec:creep} for a description of the calculations). The window had deflected approximately 75 mm. which corresponds to a radial strain of approximately 3\%, well below the typical failure strain of 10\%. In an attempt to produce a failure, we added compressed air to the system, increasing the pressure to 85 psi. At this pressure water began to leak around the gaskets, and the test was stopped. The end state of the window is shown in Figure \ref{fig:hydrostat_high_p}, with the water pressure valve showing the final pressure on the right and the water leaking out around the gaskets on the left. We note that 85 psi (586 kPa) is twice the predicted failure pressure for a safety factor of 3, a degree of uncertainty that emphasizes the importance of empirical testing of this novel structure.

Given these results, we find that the lower limit of the safety factor of the thin windows is approximately 5.7 at sea level (1 atm) and 8.2 at South Pole (0.7 atm). Because the window will never experience a pressure over one atmosphere, we can be confident that the window will not fail under normal conditions. However,  we must also explore other time-dependent factors which affect the mechanical properties of these window materials.

\subsection{Creep} \label{sec:creep}

In addition to the initial elastic deflection, plastics will undergo a visco-elastic creep deformation under continuous load. The creep deformation is logarithmic with time; thus, most of the deflection occurs very quickly, and deflection out to longer time scales is simple to predict with a model. We need to determine that the window will not creep into contact with the filters below (see Figure \ref{fig:bicep3_cutaway}). The initial elastic deflection is described reasonably well by flat plate predictions (previously described in Eq. \ref{eq:delta}) from \emph{Theory of Plates and Shells} \citep{TimoshenkoStephen1959Topa}.


We expect that the elongation of the window is azimuthally symmetric, and follows the shape of a parabola. Therefore, we can use the center deflection to predict the changed length along the arc of the parabola to find the radial strain (ie, the percent elongation). The relation is: 
\begin{equation}
\begin{split}
        \text{Radial strain} = \left(R_{\text{arc}}-R\right)/R\label{eq:rstrain1}\\
    R_{\text{arc}} = \frac{1}{2}\left[\sqrt{R^2+4\delta^2}+\left(\frac{R^2}{2\delta}\sinh^{-1}\left[{\frac{2\delta}{R}}\right]\right)\right],
\end{split}
\end{equation}
where $R$ is the initial radius and $\delta$ is the center deflection. We can relate the estimated radial strain to a power law similar to that in \cite{FindleyWilliamNichols19741cop}, obtained from observations of creep in polyethylene over sixteen years.

Figure \ref{fig:creep_def} shows the measured deflection, fit deflection, and radial strain on a thin window on the BA4 cryostat over approximately a month \citep{Petroff2024}. The radial strain is predicted to reach 4.5\% after 5 years of operation, well below the failure strains of 9--12\% (reported in Section \ref{sec:strength}). The window also remains out of contact with the foam filters below it. 

\subsection{Gas Permeation} \label{sec:perm}

As is typical in cryogenic astronomical receivers, once the cryostat is cooled, it is valved off, and the vacuum pump shut down to eliminate a source of electrical noise and vibration. The cryostat then relies on  ``cryopumping" (condensing residual gas on internal cold surfaces) to maintain vacuum through the observing season. For this reason, it is essential to ensure that the gas permeation rate through the window, which inversely scales with the window thickness, remains acceptably low.

We use both small test chamber and full receiver pressure measurements to characterize the atmospheric gas accumulation rate in a closed system. We then derive a normalized permeation rate ($Q_{\text{measured}}$) to compare between gas accumulation rates at different volumes as: 
\begin{equation} \label{eq:q_meas}
Q_{\text{measured}} = \frac{\Delta P_t}{\Delta t} \frac{V}{A}
\end{equation}
where $\Delta P_t$ is the change in pressure over the change in time $\Delta t$, $V$ is the volume of the closed system, and $A$ is the area over which permeation occurs.

These rates can also be compared to other measured atmospheric permeation rates for HDPE, such as those found in \cite{nguygen-tuong}. At a given temperature, the permeation rate depends on the permeability of a material $K$, the pressure difference between the two volumes $\Delta P_v$ (for our purposes assumed to be atmospheric pressure at a given site), the thickness ($h$) and the area over which permeation occurs ($A$):
\begin{equation} \label{eq:Q_pred}
		Q_{\text{predicted}} = \frac{K \Delta P_v}{h A}.
\end{equation}

To help establish an acceptable maximum rate of gas permeation into our cryostats through the window, we use measurements of final warm pressures from two BICEP3 seasons that had cryogenic issues associated with excess gas accumulation. The receiver has a known volume of 905 L, and has a window open diameter of 730 mm (area 418$\pow{3}$ mm$^2$). 

In 2018, a faulty o-ring seal at the window flange allowed gas to enter the cryostat over the course of one year. During the following summer season we warmed the receiver and measured the pressure inside the cryostat. The final pressure was 49 mbar after 359 days under vacuum. If all of that gas had instead permeated through the window, the window would have had a permeation rate of 3.4$\pow{-9}$ mbar L s$^{-1}$ mm$^{-2}$. The measured bolometer white noise did not increase significantly in 2018 compared to 2017, implying that this amount of gas accumulation did not adversely impact performance (see Section \ref{sec:NET} for further discussion on BICEP3's white noise performance). 

In 2022 leaks in a flange at the base again allowed excess gas entry into the cryostat. This rate of gas entry resulted in a significant increase in the measured white noise (again, see Section \ref{sec:NET} for further discussion) for approximately 463 days. However, we cannot distinguish what fraction of the gas entered during the higher noise period and what fraction entered in the previous 626 days the receiver was under vacuum. Conservatively, we average the final warm pressure of 314 mbar over the entire period the receiver was under vacuum, recognizing that it is likely that the actual problematic permeation rate was higher than this average, and posit that a permeation rate through a thin window of 7.2$\pow{-9}$ mbar L s$^{-1}$ mm$^{-2}$ will cause significant cryogenic and optical issues. For both the 2018 and 2022 cases, time at vacuum, final warm pressure and estimated upper bound window permeation rate are summarized in the bottom of Table \ref{tab:perm}.

\begin{figure}[b] 
    \centering\includegraphics[width=\linewidth]{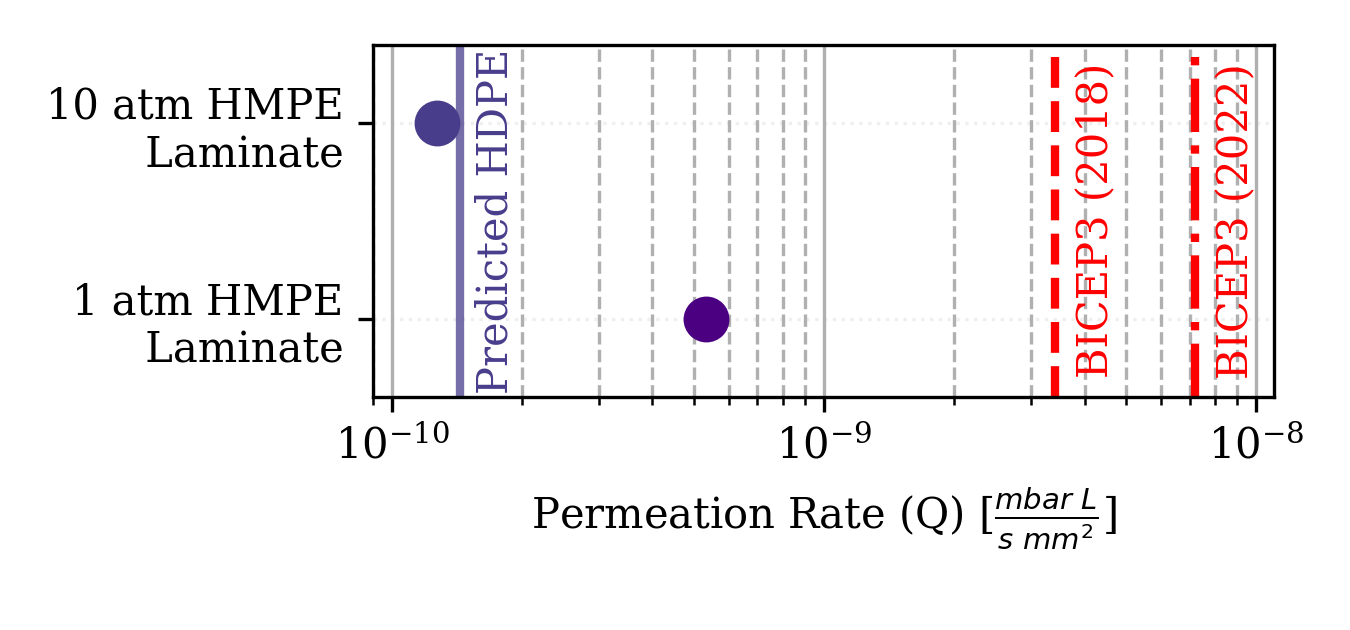}
    \caption{Measured permeation rates of small samples (circles) compared to the predicted rate with equivalently thick bulk HDPE from literature (solid line) and two limits on permeation from the unacceptable maximum rate of gas accumulation in two years; a marginally acceptable rate from BICEP3 in 2018 (dashed line) and and unacceptably high rate from BICEP3 in 2022 (dot-dashed line). The predicted HDPE value is calculated with Eq. \ref{eq:Q_pred}, the permeation constant from \cite{nguygen-tuong}, and the laminate sample thickness. The two upper limit rates are determined through final warm pressure measurements (reported in Table \ref{tab:perm}) assuming that all the gas entered through the window over the time the BICEP3 cryostat was under vacuum. Both small sample laminates are shown in Fig. \ref{fig:lam_qual}. }
    \label{fig:perm_rates}
\end{figure}

\begin{table*} 
\caption{Measured Upper Bound and Predicted Permeation Rates for Two Thin Window Cold Runs}
\begin{tabularx}{\textwidth}{*{5}{>{\centering\arraybackslash}X}}
\hline \hline
    Receiver & Time at Vacuum & Final Warm Measured Pressure & Upper Bound Permeation Rate \tablenotemark{a}& Predicted Window Permeation Rate\tablenotemark{b} \\
    & [days] & [mbar] &  [$\pow{-10}$ $\frac{\text{L mbar}}{\text{s mm}^2}$]& [$\pow{-10}$ $\frac{\text{L mbar}}{\text{s mm}^2}$]\\
    \midrule
    BA4  & \multirow{2}{*}{30} & \multirow{2}{*}{0.85} & \multirow{2}{*}{7.1} & \multirow{2}{*}{1.44} \\
    (Sea Level) &  &  &  &  \\
    BA150 & \multirow{2}{*}{341} & \multirow{2}{*}{2.1} & \multirow{2}{*}{1.5} & \multirow{2}{*}{1.21} \\
    (South Pole) &  &  &  &  \\
    \midrule
    \multicolumn{5}{c}{Measured Upper Bound Permeation Rates Associated with Cryogenic Issues} \\
    \hline \hline
    BICEP3 in 2018 & \multirow{2}{*}{359} & \multirow{2}{*}{49} & \multirow{2}{*}{34} & \multirow{2}{*}{} \\
    (South Pole) &  &  &  &  \\
    BICEP3 in 2022 & \multirow{2}{*}{1089} & \multirow{2}{*}{314} & \multirow{2}{*}{72} & \multirow{2}{*}{} \\
    (South Pole) &  &  &  &  \\
    \midrule
\end{tabularx}
\tablenotetext{a}{Estimated upper bound permeation rate through a window using Eq. \ref{eq:q_meas}, the reported time at vacuum, and final measured warm pressure, assuming that all receivers have the same internal volume (905 L) and window open diameter of 730 mm (area 418$\pow{3}$ mm$^2$).}
\tablenotetext{b}{Predicted permeation rates of the thin windows on the receiver using Eq. \ref{eq:Q_pred} and permeation constant from literature \citep{nguygen-tuong}.}
\label{tab:perm}
\end{table*}
To isolate the laminate windows' permeation rate from other sources of gas accumulation in a vacuum, we tested two small scale samples (shown in the top of Figure \ref{fig:lam_qual}) on a small vacuum vessel over a few days. Previously, we had noted an unacceptably high permeation rate in thin windows that were laminated at one atmosphere of pressure \citep{Barkats2018,Eiben2022}. Therefore, we also compared the permeation rates of low pressure laminates and high pressure laminates, to confirm that higher pressure lamination reduced the permeation rate. Figure \ref{fig:perm_rates} shows the results of the small scale permeation rate testing. The measured small scale permeation rates are compared to the predicted permeation rate for bulk HDPE of equivalent thickness (1.4 mm) using Equation \ref{eq:Q_pred} with the permeation constant found in \cite{nguygen-tuong}. 

Both the measured and calculated window permeation rates in Figure \ref{fig:perm_rates} are significantly lower than the previously discussed problematic permeation rates estimated from two years with cryogenic issues on BICEP3 (red dashed and dot-dashed lines). However, the 1 atm laminate has a rate over three times higher than the 10 atm laminate; this is likely due to low laminate fill factor at low lamination pressure. The laminating LDPE is more effectively forced around the HMPE fibers at high pressure, resulting in better laminate consolidation and homogeneity. We will continue to use Eq. \ref{eq:Q_pred} and the \cite{nguygen-tuong} permeation constant to predict permeation rates of windows, given the measured consistency between the 10 atm laminate rate and the predicted rate.

We have since run an observing season with a high pressure laminate window on a cryostat (BA150) at the South Pole (at altitude, with assumed 0.7 atm pressure difference) and warmed the receiver. We also measured the gas accumulation in another receiver (BA4) at sea level, though the run was over a shorter period of time \citep{Petroff2024}. Table \ref{tab:perm} reports the time at vacuum and final pressures for each run. The final warm pressure can be used to estimate an upper bound permeation rate through a thin window, which can be compared to the predicted permeation rate for a given window, or the upper bound window permeation rates associated with cryogenic issues.

Both of the receivers with thin windows cooled and remained cold successfully. The predicted window permeation rates are below the upper bound rate estimated from the total gas accumulation, indicating that the permeation through the windows is a significant contributor to gas accumulation inside the cryostat. The window likely dominates over other sources of gas accumulation (such as permeation through the o-rings) over long timescales. These measured amounts of gas are also significantly below the previous problematic gas amounts. Roughly, we may estimate that the thin windows may allow a cryogenically problematic amount of gas into the cryostat after about twenty years, wherein the gas could be removed by warming, pumping out, and cooling again. The BICEP/\emph{Keck} collaboration has very rarely kept a cryostat cold for longer than three consecutive years; permeation of gas through the window into the cryostat is unlikely to become a problem.

\section{Optical Tests}\label{sec:optical_tests}
The vacuum window is the single highest contributor to optical noise in the instrument. As such, characterizing and minimizing the optical impact of the window on our instrument is critical. After mechanical considerations are satisfied, we design the window to minimize reflected power across our observing band (Section \ref{sec:antiref}), and characterize the potential polarizing effect of the laminates at our frequencies (Section \ref{sec:polarization}). We also report in-situ characterization on the BICEP3 telescope, via direct optical measurements between different window materials with the CMB detectors (Section \ref{sec:BICEP3_char}).

\subsection{Anti-reflection} \label{sec:antiref}
\begin{figure*}[t] 
    \centering\includegraphics[width=15cm]{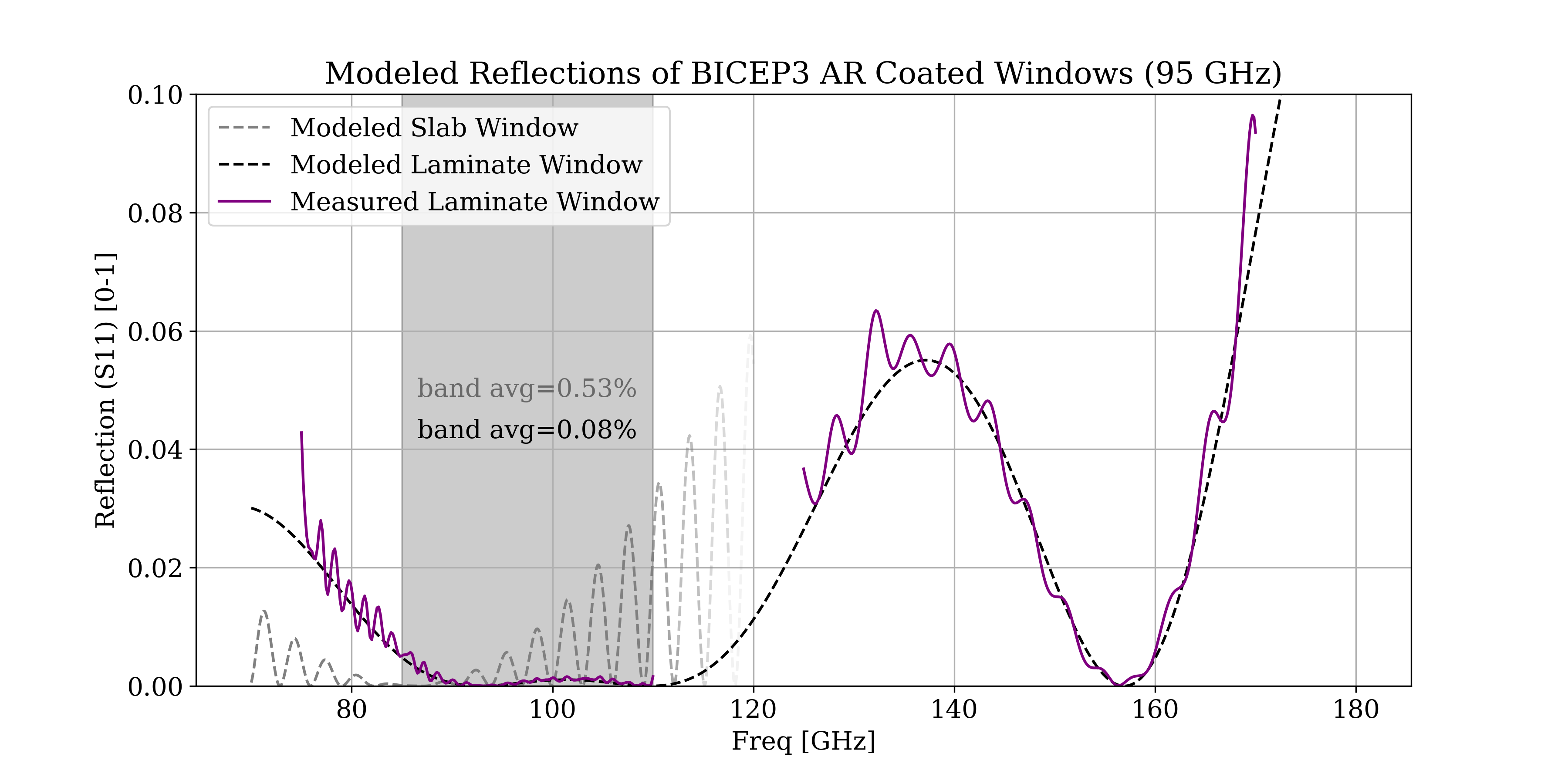}
    \caption{Measured VNA $S_{11}$ reflections of the BICEP3 thin window (solid purple lines, both in-band WR 10 and out of band WR 6.5), and modeled reflections of the laminate (dashed black line) and the slab window (dashed gray line). The average in-band reflection of the AR coated slab is estimated to be 0.53\% while the average in-band reflection of the AR coated laminate is estimated to be 0.08\%. The in-band and out-of-band measurements on the laminate window show excellent consistency with the modeled reflection.}
    \label{fig:reflect}
\end{figure*}
Like all of the optics in the system, the window surfaces require additional anti-reflection (AR) layers to minimize their reflectivity. These layers have carefully selected indexes of refraction and thicknesses to cause destructive interference of reflections within our observing band. This is needed to minimize signal loss and to suppress ghosting and interference artifacts caused by reflections within the receiver. We have developed a process that allows us to fabricate and laminate matching layers with accurately tuned refractive index as well as thickness \citep{Eiben2022, Dierickx2021}. In addition, the thin laminated window can be made resonantly transmissive over a wide band by choosing its thickness to be a small integral number of half-wavelengths at the band center. This has enabled us to fabricate a complete window assembly with extremely low in-band reflection, using simple single-layer quarter-wave coatings.

We have taken validation measurements of the deployed AR coats for BICEP3 (shown in Figure \ref{fig:reflect}) and BA150 (discussed in Table \ref{tab:ar_prop}). In Table \ref{tab:ar_prop}, core thicknesses were measured with calipers, and the modeled values for the AR parameters are derived from vector network analyzer (VNA) measurements, such as those in Figure \ref{fig:reflect}.

We measured the millimeter-wave reflectivity of AR coated windows using a VNA equipped with WR-10 (75 to 110 GHz) and WR-6.5 (110 to 170 GHz) frequency extenders in a single port configuration ($S_{11}$). Each waveguide frequency extension VNA head has a rectangular 23 dBi standard gain horn and 90° offset parabolic mirror to produce a Gaussian beam waist at the sample position.  Samples are placed at the Gaussian beam waist perpendicular to the beam axis. The reflection from the sample is normalized to that from a machined aluminum plate, which is assumed to have near perfect reflectivity. Time gating of the VNA reflection signal is used to eliminate spurious reflection signals from the horn, collimating mirror, and beam termination. This measurement setup was also described and shown in \cite{Eiben2022,Eiben2024}. 

We can fit a reflection model to these measurements; given we have a good constraint on the physical thickness from mechanical measurement, the fit is primarily to estimate the index of refraction for either the core or the core and AR coat, depending on the sample. The index of any given material is related to that material's density, and therefore in the case of a laminate composite, the composite fill factor. In \cite{Barkats2018}, the low fitted value of the index of refraction for laminate windows (n$\approx$1.35) was attributed to a low fill factor of 30--40\%: this is partially true, but those index measurements were later found to be affected by systematic error. With proper systematic correction, we still find reduced index (n$\approx$1.45) for low pressure (1 atm) laminates due to low fill factor, but we now estimate the fill factor of 1 atm laminates to be approximately 85\%. Fully consolidated high-pressure laminated windows are found to have indexes similar to that of bulk polyethylenes, between 1.536$<$n$<$1.557 \citep{Lamb1996,delessandro2018,Elwood2024}; the range of indexes is due to repeatability in the $S_{11}$ measurements. These fill-factor from index estimates are also corroborated with density measurements. 
\begin{table*}
\caption{Design, Measured and Modeled AR Coat Parameters for Deployed Thin Windows}
\begin{tabularx}{\textwidth}{*{7}{>{\centering\arraybackslash}X}}
    \hline \hline
    \multirow{2}{*}{Receiver} & \multicolumn{2}{c}{Core Thickness [mm]} & \multicolumn{2}{c}{AR Thickness [mm]} & \multicolumn{2}{c}{AR Index} \\
    \cmidrule(r){2-3}\cmidrule(r){4-5}\cmidrule(r){6-7}
     & Design & Measured\tablenotemark{a}& Design & Modeled\tablenotemark{b}& Design & Modeled\tablenotemark{b}\\
    \midrule
    BICEP3 & 1.45 & 1.38 & 0.63 & 0.61 & 1.25 & 1.25\\
    BA150 & 1.20 & 1.18 & 0.39 & 0.40 & 1.23 & 1.23\\
\midrule
\end{tabularx}
\tablenotetext{a}{Measured with calipers.}
\tablenotetext{b}{Modeled and fit to VNA measurements, such as those in Fig. \ref{fig:reflect}.}
\label{tab:ar_prop}
\end{table*}
The AR coat deployed on the previous bulk polyethylene windows were generated before AR process improvements described in \cite{Eiben2022} and \cite{Dierickx2021}, and therefore the designed null is off-center in the band, as seen in Figure \ref{fig:reflect}. Additionally, the thin window allows for a second designed reflection null in-band, and eliminates the reflection fringes that the thick slab produces. The very thin nature of the laminate windows allows for exceptionally low reflection with a single layer AR coat. However, we must laminate at high pressure (above 4 atm) to generate a laminate window with the expected index, which is vital for designing AR coats. In the next section we discuss an additional potential optical issue with poor lamination, which may affect the polarized transfer function of the receiver. 

\subsection{Birefringence} \label{sec:polarization}
The BICEP/\emph{Keck} Collaboration is aiming to measure the polarization of a cosmological signal to very high precision. The B-mode angular power spectra are expected to be multiple orders of magnitude below the temperature anisotropies. Due to the highly aligned structure of the HMPE fibers within the woven fabric at macroscopic scales, there is a legitimate concern that the woven structure we are incorporating into these windows may induce a polarized response, reducing our sensitivity to primordial polarization states. An instrumental polarized effect, such as birefringence, from the window would induce correlations that our normal analysis self-calibrates for and eliminates \citep{Keating_selfcal,BK18}. Nevertheless, we constrain any potential polarizing effect from the laminate window using both the benchtop measurements described in this section and direct measurements on the BICEP3 receiver in Section \ref{sec:t_to_p}.

\begin{figure*}[t]
    \centering
    \includegraphics[width=0.53\linewidth]{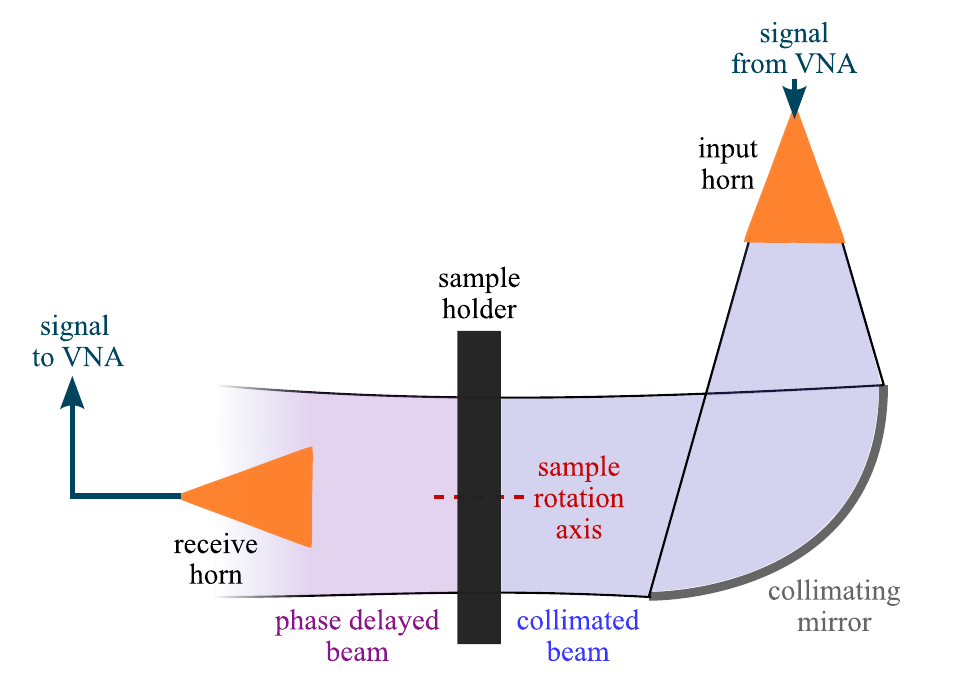}
    \includegraphics[width=0.4\linewidth]{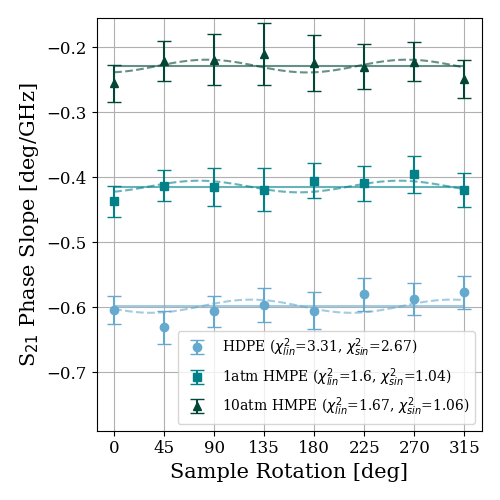}
    \caption{[Left] Diagram of phase offset measurement (not to scale). Millimeter wave signal comes from the VNA and is coupled to free space through the input horn and collimated by the parabolic mirror. The collimated beam then interacts with the sample in the sample holder orthogonal to the propagation axis, where the phase will be delayed by the sample's electrical length. The receive only horn then couples the signal back into the VNA ($S_{21}$ phase measurement). [Right] Fit phase slopes for S$_{21}$ phase offset measurements of 0.79 mm (1/32") thick HDPE and the two HMPE laminates, one baked at 1 atm (0.68 mm thick) and the other baked at 10 atm (0.4 mm thick). If the materials are not birefringent then the phase slope should not vary significantly with sample rotation. If the materials exhibit uniaxial birefringence, we would expect the same phase slope on measurements 180 deg rotated from each other: we have fit both a flat line (solid lines, $\chi_{lin}^2$ in the legend) and a sin($2\theta$) wave (dashed lines, $\chi_{sin}^2$ in the legend) to the data to test the maximum amplitude of the possible phase offsets for each material.}
    \label{fig:el_len}
\end{figure*}

The Stokes Q (linear horizontal or vertical polarization), U (linear $\pm$ 45 deg polarization), and V (circular polarization) are a useful basis to define in this context. Given that our orthogonal linear antennas couple only with the Q polarization, our detectors should not see any U signal \citep{Keck2015B}. This assumes that our linear antennas are perfectly orthogonal, and that our instrument does not rotate the polarization of the input signal. Birefringent optical materials have a ``slow" and ``fast" axis, which will retard the phase of incident polarized light, potentially transforming a hypothetical pure Q signal into some other combination of these Stokes parameters. Birefringence depends on the difference in index and path length (i.e., electrical length) between the slow and fast axis, and the orientation of the axes to the incident polarized light. There are two potential impacts that a birefringent window could have on our receiver's polarized transfer function:

\begin{enumerate}
    \item Rotation of polarization, or introducing a U beam.
    \item Reduced polarization efficiency, or rotation from Q partially into U and V.
\end{enumerate}

If there is a birefringent effect from the difference in index between the HMPE fibers and the surrounding LDPE, the windows are thin enough that there should not be a long enough path to produce a significant phase offset. However, it is also possible that the HMPE weave induces other polarization effects, such as polarized scatter, that in practice act like a birefringent material. Additionally, as discussed previously in Section \ref{sec:antiref}, the lamination quality (fill factor) has an effect on the index of the laminate. The lower pressure laminates have a lower fill factor and therefore lower index; the HMPE weave in lower pressure laminates is even more visible due to the poor lamination (see Figure \ref{fig:lam_qual}). If birefringence or other polarized effects were to occur, it is more likely to be measurable in a low pressure laminate where gaps between the HMPE and LDPE are more prevalent.

However, multi-layer laminates---such as the deployed windows---have three layers of HMPE weave rotated 30 degrees offset from each other, as discussed in Section \ref{sec:manufact}. Multiple weave orientations may obscure the relation between the weave and the phase offset. To better understand the polarization properties of the laminated fibers, we test single layer HMPE laminates to reduce potentially confounding additional angles.

We can compare between a poorly laminated (1 atm) single layer laminate, a well laminated (10 atm) single layer laminate, and a thin extruded bulk HDPE sample. Extruded bulk HDPE windows have been observed to exhibit birefringent properties at far IR frequencies, so we use it as a detection upper limit \citep{Paine2013}. To measure the transmission ($S_{21}$) phase we used the previously described $S_{11}$ VNA set up with an additional receive-only port using an even harmonic mixer to downconvert the received signal. This setup is shown in the left of Figure \ref{fig:el_len}. When we introduce a material to the beam of our VNA, it will introduce a phase offset due to the change in electrical length of the path between the horns that varies as:
\begin{equation}
    \Delta\phi = \frac{2\pi n h}{\lambda} - \frac{2\pi h}{\lambda} ,
\end{equation}
where $\Delta\phi$ is the change in phase (in radians) referenced against the same length without the material ($n=1.0$), $n$ is the index of refraction, $h$ is the thickness of the material, and $\lambda$ is the wavelength of the incident light. The electrical length is the product of the index and the thickness of the material. Therefore, changes in the index at different orientations will result in a different offset of phase, and a different electrical length.

We centered a small rotating stage between the horns, such that the rotation axis was centered with the beam, then took phase measurements on 45 degree increments. We then fit the slope of the phase offset over frequency (fit slopes for each rotation measurement and material shown in the right Figure \ref{fig:el_len}), which is directly related to electrical length. The repeatability of the phase was 3 degrees: this is very similar to the calculated fit uncertainties shown in at the right of Figure \ref{fig:el_len}.

If the materials are not birefringent, then we expect no relationship between the sample rotation angle and the fit phase slope. If the material exhibits uniaxial birefringence (one fast axis and one slow axis orthogonal to each other), then we expect the same change in electrical length on measurements rotated 180 degrees from each other; in other words, $S_{21}$ phase slopes fitting a sine wave with a fit phase offset ($x$), amplitude ($A$), and amplitude offset ($y$) over the sample rotation ($2\theta$):
\begin{equation}
    \Delta\phi = Asin(2\theta + x) + y 
\end{equation}
We fit both a flat line and a sin(2$\theta$) wave to the measurements to test the null hypothesis. The sine wave fit has two more degrees of freedom than the flat line, and therefore the $\delta \chi^2$ between the flat fits and the sin(2$\theta$) fits would have to exceed the extra degrees of freedom to be significant. The $\delta \chi^2$ between each materials' fits are approximately 0.6 (specifically, the HDPE $\delta \chi^2=0.64$, the 1 atm HMPE $\delta \chi^2= 0.56$ and the 10 atm HMPE $\delta \chi^2=0.60$)---well below the two extra degrees of freedom in the birefringent sin(2$\theta$) model---which does not allow us to reject the null hypothesis. 

We might expect more than one fast or slow axis induced by the HMPE weave in the laminate. However, these measurements cannot constrain more than a uniaxial birefringent effect. We would also expect that extruded HDPE would act more like a traditional uniaxial birefringent material---as we suspect that the extrusion should induce a preferred orientation of polymer crystals along one axis in the manufacturing process---which should be captured by these measurements \citep{Paine2013}. The HDPE was also the thickest of the samples tested, and therefore would induce the largest phase difference and the greatest chance of detection. There is no strong evidence for a detection of birefringence in the extruded HDPE sample. As such, we set an upper limit on birefringence in the laminate windows at the maximum level of the fit sin(2$\theta$) amplitudes. The maximum fit amplitude in electrical length is approximately 0.008 mm for any of the materials, or approximately a 0.02 change in index for the thinnest laminate.

An upper limit change of 0.02 in the index is supported by the typical repeatability of $S_{11}$ index of refraction fits to laminate windows (fit indexes range between 1.536 to 1.557, as discussed in Section \ref{sec:antiref}). If we calculate the impact of such a birefringent window on a fully polarized incident Q beam, we can put constraints on those potential impacts. We can take the combination of window parameters that will produce the worst phase retardation: the thickest deployed window (1.4 mm on BICEP3), and the highest frequency a deployed window is designed for (170 GHz, the upper edge of the BA150 band) to constrain the worst case scenario. The window can be modeled with a Mueller matrix such as:
\begin{equation} \label{eq:mueller}
    \begin{split}
        M_{\text{window}}(\phi) &=  
            \begin{bmatrix} 1 & 0 & 0 & 0 \\ 
                            0 & 1 & 0 & 0 \\ 
                            0 & 0 & -\cos{\phi} & -\sin{\phi} \\ 
                            0 & 0 & \sin{\phi} & \cos{\phi} 
            \end{bmatrix} \\
        \phi &= 2 \pi (n_s - n_f)h/\lambda,
    \end{split}
\end{equation}
where $\phi$ is set by our worst case parameters, the thickness ($h$) of 1.4 mm, the fast and slow axis indexes are set have a 0.02 difference ($n_f$ = 1.536 and $n_s$ = 1.556), and $\lambda$ is the wavelength of interest. This matrix will rotate depending on the eigenstate (orientation) of the birefringent window relative to the input polarization. We then multiply the matrix by an input Stokes vector (for our purposes, purely +Q) to generate an output Stokes vector of polarization states transformed by the hypothetically birefringent window.

\begin{table}
    \caption{Stokes Parameters thru Worst Case Laminate\tablenotemark{a}}
    \begin{tabularx}{0.47\textwidth}{*{3}{>{\centering\arraybackslash}X}}
    \hline \hline
    Stokes & 22.5 deg & 45 deg \\
    \midrule
    Q (S1) & 0.999 & 0.996 \\
    U (S2) & 0.002 & 0.000 \\
    V (S3) & 0.066 & 0.095 \\
    \midrule
    \end{tabularx}
    \label{tab:birefrin_stokes}
    \tablenotetext{a}{Worst case laminate parameters are a combination of the thickest deployed window (1.4 mm) at highest deployed frequency (170 GHz) oriented at the worst eigenstates (22.5 and 45 deg).}
\end{table}

There are two eigenstates (orientations of the window relative to the linear polarization) that produce the worst rotation for our purposes. At 22.5 degrees the window will rotate the incident Q beam the furthest into U, and at 45 degrees the Q beam will be the furthest rotated into V. The calculated rotations of a hypothetical birefringent window are shown in Table \ref{tab:birefrin_stokes} at these two worst orientations. It is highly unlikely that the windows are oriented at the worst eigenstates, but these calculations can be taken as the upper limits. The maximum power possibly transformed into a U beam is 0.2\%, significantly less than the previously constrained BICEP2 and Keck Array U beams with powers of 0.8\% \citep{Keck2015B}. This effect should be a uniform rotation of the monopole, which would be taken out during the beam fitting process.

The maximum reduction of polarization efficiency is estimated by the difference of the full intensity to and the output Q parameter:
\begin{equation}
    \epsilon = \sqrt{Q^2 + U^2 + V^2} - Q = 1 - Q.
\end{equation}


Which we can see remains very low (0.3\%), roughly half of the previously constrained BICEP3 polarization efficiency \citep{bicep3}. Fully characterizing the polarization efficiency of a receiver requires a full end-to-end polarization measurement, which is very difficult to accurately conduct \citep{Cornelison2022,Cornelison2024}. We plan to run another absolute polarization calibration on BICEP3 with a thin laminate window in an upcoming summer season, and the relative change in polarization angle of detectors will then be directly compared.

Though a hypothetically birefringent window results in negligibly uniform rotations of polarized light to first order, a higher order change in the polarized transfer function of the telescope would not be well characterized by these estimates. In Section \ref{sec:t_to_p} we explore the effects of how the thin window changed the residual beams in BICEP3, and therefore the estimated temperature to polarization leakage.

\section{Optical Characterization on BICEP3} \label{sec:BICEP3_char}
To explore in more detail the performance of these thin windows, we sent them to the South Pole in the austral summer of 2022--2023 as the deployed science-grade window on the BA150 receiver, and to replace the existing window on BICEP3. This enabled us to test the windows optically on their respective receivers, both through dedicated calibration measurements before installation and through regular observations. Of particular interest was the impact of a thin window on BICEP3, the 95 GHz receiver that has been observing at the South Pole since 2016. Because that receiver has operated previously with a slab HDPE window in 2017 and 2018 and an UHMWPE window for 2019 through 2022 we can directly compare each of the polyethylene windows on the same instrument. 

There were three major sources of concern regarding the impact of this new window on the optical performance of BICEP3. The polarized residual beam response, also known as temperature-to-polarization leakage, is discussed in Section \ref{sec:t_to_p}. The optical load from the window seen by the detectors is discussed in Section \ref{sec:opt_load}. Finally, the measured change in the detector white noise within the receiver is discussed in Section \ref{sec:NET}.

\subsection{Temperature-to-Polarization Leakage}\label{sec:t_to_p}
As discussed in Section \ref{sec:polarization}, the polarized transfer function of the telescope may be changed by the thin window. Since we carry out polarimetry by differencing co-located, orthogonally-polarized detectors, any deviations in beam response between orthogonal pairs results in differential beam response. By doing a Taylor expansion of the differential beam, we can predict the shape of the differential beam response for the lowest order modes (up to order 2). We then regress our maps against a known temperature sky (i.e., a Planck temperature map) to fit the amplitude of these modes, and deproject their response \citep{PlanckNPIPE}. The residual beam response after deprojection is responsible for temperature-to-polarization (T-to-P) leakage, as the undeprojected beam response leaks the bright temperature sky into our polarization maps. This is currently our main source of systematic uncertainty \citep{beams2019,BK18}.

To test for T-to-P leakage we must first characterize the beam shape of our detectors through far-field beam mapping. During the austral summers of 2016 through 2019 and 2023, we raster-scanned a chopped thermal source located on a tower approximately 200 m away in order to map the far-field response of BICEP3. During a far-field beam mapping campaign, we carry out 40--50 of these raster scans, between which we rotate the telescope in boresight to ensure full beam coverage \citep{bicep3}. After a campaign, we combine the beam maps into stacked, high-fidelity ``composite" beam maps.

For each schedule, we fit a seven parameter elliptical Gaussian to the beam map of each detector \citep{beams2019}. Comparing the per detector fitted beam parameters between 2019 and 2023, the only significant change in beam parameters was a systematic decrease in beamwidth of 1.2\% \citep{Giannakopoulos2024}. We might expect this if the slab window acts like a meniscus lens on the beam, with two different curvatures on the top and bottom interfaces as the window deflects. The thin window therefore reduces the impact of the window on the focus of the instrument, reducing the measured beamwidth.

To assess the potential impact of this T-to-P leakage, we take the composite beam maps and simulate observing a pure temperature sky. These simulated measurements are then run through the rest of the analysis pipeline: any resultant power in the polarized maps must have been induced by undeprojected beam mismatch during the simulated observation. We checked the simulated temperature-to-polarization power spectra for every year that we had independent beam measurements (2016 through 2019 and 2023), and found no significant increase in the simulated BB power signal in 2023 with a thin window on BICEP3. Since T-to-P leakage is the dominant instrumental systematic of our telescopes, we therefore expect that the thin window does not significantly impact our B-mode measurements \citep{BK18}.

\subsection{Measured Optical Load}\label{sec:opt_load}
\begin{figure*}[ht!]
    \centering
    \includegraphics[width=0.45\linewidth]{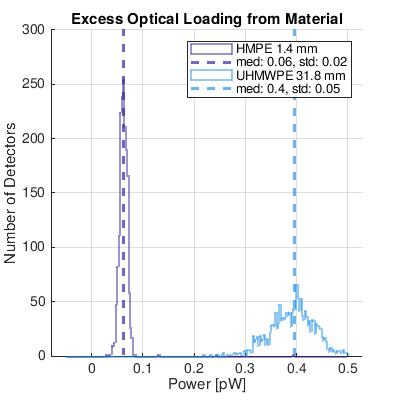}
    \includegraphics[width=0.45\linewidth]{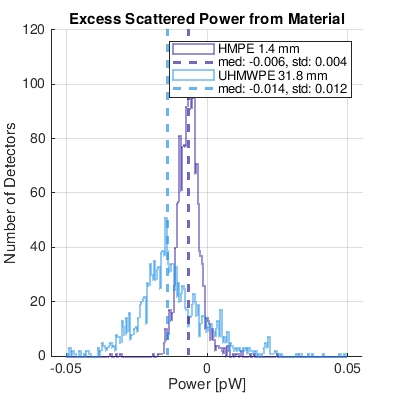}
    \caption{[Left] Histograms of excess loading per detector from two different windows placed above BICEP3. [Right] Histograms of scattered power to the forebaffle associated with placing the material in front of the BICEP3 receiver. Blue line is a spare UHMWPE 31.8 mm (1.25") thick window for BICEP3; dashed line denotes the median excess load from the slab. Purple line is the replacement HMPE 1.4 mm thick window for BICEP3; dashed line is the median excess load from the laminate.}
    \label{fig:radiometric}
\end{figure*}
The anticipated decrease in in-band instrumental optical load from decreasing the window thickness is a primary driver for developing thin laminate windows, as shown in Figure \ref{fig:intro_load_net}. We can test the difference in optical load with our receivers themselves, as they are highly sensitive and offer the most direct measure of receiver response. We call such measurements ``radiometric," as they directly measure the change in optical load with the receiver.

Our detectors are transition edge sensors (TES), which voltage-bias a superconducting element---either titanium or aluminum depending on the measurement---on the superconducting transition edge. We measure optical power variation by compensating with electrical power. A given detector will have a designed total saturation power, and changes to the applied optical power will change the margin to that saturation power \citep{bicep3}. To gain a direct measure of the power on a detector we can vary the applied bias voltage out of the transition, then step down in voltage until the detector is superconducting. These measurements are known as partial load curves (PLCs); partial because we only explore a range around the predetermined bias voltage. PLCs allow us to quickly probe the difference in optical power between two instrument states, such as adding and removing an optical component in front of the receiver, assuming that there is no other change to the applied optical or electrical power. If one is careful in taking a variety of reference measurements, one can attempt to determine not only the optical loading from transmission loss in the material, but also any additional optical loading from scattered power associated with the material.

First, we take reference measurements of the receiver without a material ($R$). We subsequently add the absorptive forebaffles to the receiver and perform the same measurement ($R_{fb}$). Each of these two measurements are repeated with a material placed on top of the receiver ($M, M_{fb}$). The excess load from the material itself and the excess scattered power associated with the added material can therefore be derived by ($M-R$) and [\((M_{fb} - M) - (R_{fb} - R)\)] respectively. Histograms of measured powers per detector for the BICEP3 HMPE window and a UHMWPE spare slab window are reported in Figure \ref{fig:radiometric}.

As expected, the thin 1.4 mm HMPE window causes significantly less optical loading on the detectors compared to the 31.8 mm (1.25") UHMWPE window. However, the magnitude of the difference is not as large as expected from the assumption that the polyethylene absorption losses are the same: if the loss tangent is the same in both polyethylene windows the difference in thickness between windows would result in a factor of 20 difference in optical power. With the radiometric measurements we only measure a factor of 6.6 difference in optical loading. We therefore conclude that the two polyethylenes likely have different loss tangents. This is supported by the differences in noise equivalent temperature (NET) between years, as discussed in Section \ref{sec:NET}. 

Using an optical model of the BICEP3 instrument (based on the optics described in the BICEP3 instrumentation paper and using the BoloCalc code \citep{bicep3,BoloCalc}), we can use the measured excess load from the material to estimate the $\tan\delta$ of the two materials. Having done this, in conjunction with the NET models shown in Figure \ref{fig:B3_net_loss_diff}, we estimate that the loss tangent of the UHMWPE is approximately 1.2$\pow{-4}$, and the HMPE loss is approximately 2.4$\pow{-4}$ (shown in Figure \ref{fig:intro_load_net}). Independent  laboratory measurements of the same material in \cite{Elwood2024} have obtained a UHMWPE $\tan\delta$  of (1.49 $\pm$ 0.28)$\pow{-4}$, consistent with these results.

As previously described in Section \ref{sec:design}, the forebaffle (shown in Figure \ref{fig:bicep3_cutaway}) terminates scattered light exiting the receiver on an absorber of known temperature. By taking PLC measurements with the materials with and without the forebaffles on the receiver, we can directly measure the excess in optical load associated with the scattered light through these windows. Shown in the right side of Figure \ref{fig:radiometric} the measured scattered power associated with the window materials is very small, and for both materials within two standard deviations of zero. These measurements put upper limits on the scattered fraction from these materials, at roughly 4\% for the UHMWPE and 10\% for the HMPE laminate. The amplitude of these upper limits on scattered power from window material have been corroborated by benchtop scatterometer measurements reported in \cite{CorriganLiam2019Doam}.

These radiometric  measurements provide a powerful cross check of the excess load from warm optics in front of the receiver. We can use this information as a supplement for optical modeling, and to provide estimates of optical material properties, such as in Figures \ref{fig:intro_load_net} and \ref{fig:B3_net_loss_diff}. Independent measurements of material optical load are particularly useful for estimating the receiver's photon load, which is the dominant contributor to the total per-detector white noise, discussed in the next section.

\subsection{Noise Equivalent Temperature} \label{sec:NET}
There are three major noise sources in a CMB telescope: readout noise (dependent on readout electronics), phonon noise (dependent on detector properties), and the dominant photon noise (emission from astrophysical sources, the atmosphere, and the telescope). As previously discussed, a primary driver for decreasing the thickness of the vacuum window is to reduce the optical load---and therefore photon noise---on the detectors, as we expect that the window is the dominant source of instrument photon emission. Decreasing the noise on the detectors results in a direct improvement on mapping speed.

\begin{figure}[t!]
    \centering\includegraphics[width=\linewidth]{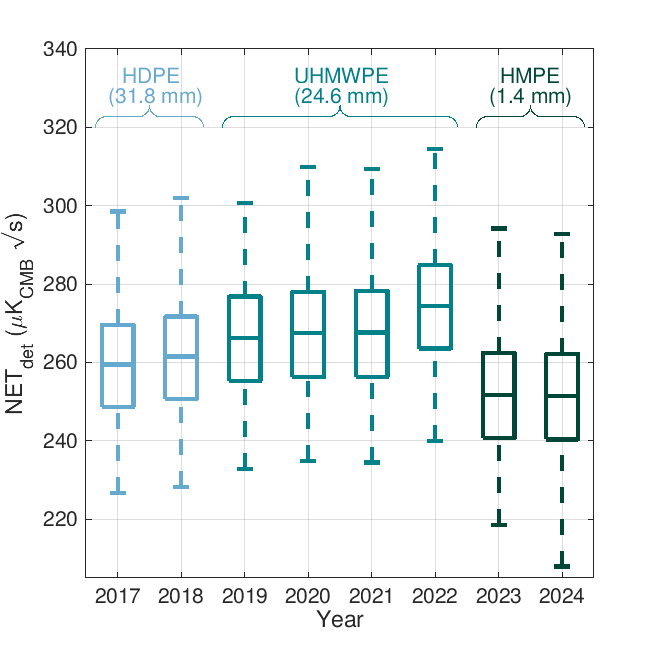}
    \caption{BICEP3 median white noise (NET) distribution over the same population of detectors per year between 2017 and 2024. Per detector per scanset (roughly 50 minutes of time ordered data) NETs are calculated, then a median per detector over the entire year's scansets are shown, with 25th (q$_{25}$), 50th, and 75th (q$_{75}$) percentiles shown in the box, to demonstrate average noise of the receiver between years. Whiskers cover the remainder of the distribution, excluding outliers, defined by q$_{25/75}$ $\pm$ (q$_{75} - $q$_{25}$). Color corresponds to the kind of window on the receiver: light blue is HDPE 31.8 mm (1.25"), blue is UHMWPE 24.6 mm (0.97"), and dark green is HMPE 1.4 mm. There is a significant change in noise between three observing seasons: 2018--2019, 2021--2022, and 2022--2023.}
    \label{fig:B3_allyear_NET}
\end{figure}

We process the raw CMB timestreams to extract a median metric of the yearly noise level for all detectors, which is calibrated into a noise equivalent temperature (NET) using publicly available Planck temperature maps \citep{bicep3,PlanckNPIPE}. Within the focal plane there are detector parameter variations, which result in a consistent distribution of measured NETs over the detector population. We compute a median over all detectors to get a representative per-season number for the entire system. Optical changes between years result in a coherent offset of the median yearly NET. If we look at the median NET per year over the last seven years of data collection on BICEP3, we can see how the NET has changed year to year, as shown in Figure \ref{fig:B3_allyear_NET}. There are three occasions when the NET change between years was greater than 1.5\%: between 2018 and 2019, 2021 and 2022, and 2022 and 2023.
From statistical modeling we expect that the yearly absolute calibration uncertainty for BICEP3 remains below 0.5\% ($\sim$1 $\mu K \sqrt{s}$ noise), which does make the greater than 1.5\% differences we are seeing between those three seasons significant. 

All three of these significant changes in noise are correlated with optics changes. These changes were:

\begin{enumerate}
    \item Between 2018 and 2019: A window change. The window thickness decreased (31.8 mm to 24.6 mm), and the material changed (HDPE to UHMWPE). The median per detector yearly NET increased 1.9\% (262 $\mu K \sqrt{s}$ to 267 $\mu K \sqrt{s}$).
    \item Between 2021 and 2022: High optics temperatures. A leaky flange led to many cryogenic optics' temperatures to be elevated. The median yearly NET increased 3.0\% (267 $\mu K \sqrt{s}$ to 275 $\mu K \sqrt{s}$).
    \item Between 2022 and 2023: A window change. The window thickness decreased (24.6 mm to 1.4 mm), and the material changed (UHMWPE to HMPE laminate). The median per detector yearly NET decreased 9.1\% (275 $\mu K \sqrt{s}$ to 252 $\mu K \sqrt{s}$).
\end{enumerate}

\begin{figure}[t!]
    \centering
    \includegraphics[width=\linewidth]{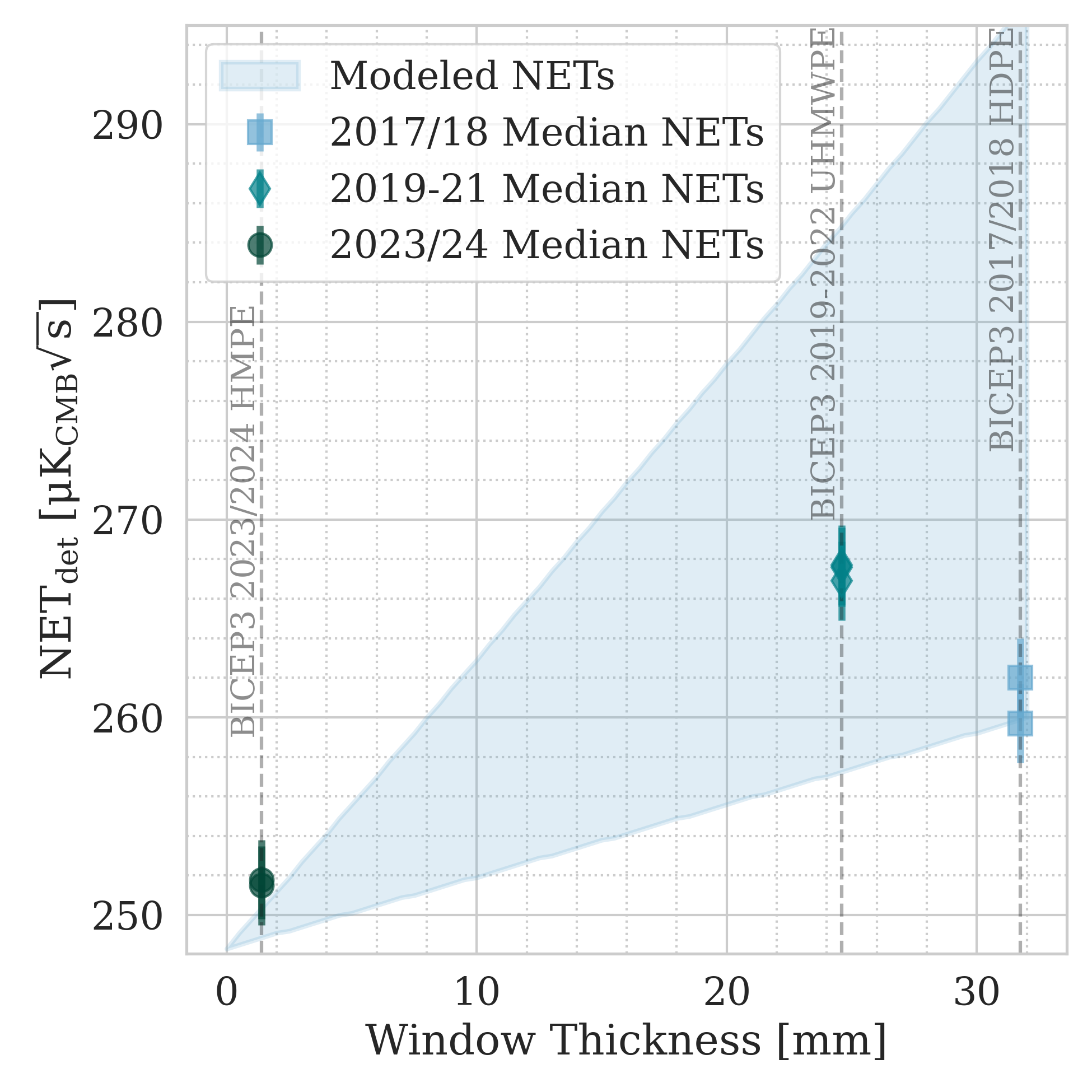}
    \caption{Median measured NETs (white noise) from Figure \ref{fig:B3_allyear_NET} as a function of the window thickness for that respective year. We exclude 2022 due to elevated optics temperatures. The blue shaded band of NET models show predicted BICEP3 NET values, varying the window loss tangent between 0.6$\pow{-4}$ to 2.4$\pow{-4}$. We have not yet concluded on the cause of the increase in the white noise measured between the thicker HDPE window (31.8 mm) in 2018 to the UHMWPE window (25.4 mm) in 2019: both a difference in material loss and a difference in scattering properties could explain this change. The HMPE window did definitively and significantly decrease the measured white noise compared to both thick slab windows.}
    \label{fig:B3_net_loss_diff}
\end{figure}

We hypothesize that each of these changes in the noise of the instrument are due to these optical changes, and that the overall decrease in noise with the thin window is likely due to the thin window alone. To validate this, we built an optical model to predict the effect that decreasing the thickness of the window has on both the optical load and white noise using the radiometric data reported in Figure \ref{fig:radiometric} and the median yearly NETs reported in Figure \ref{fig:B3_allyear_NET}. We use median detector parameters reported in \cite{bicep3} for this model, and it is used to predict how varying the band center changes NET and optical load changes in Figure \ref{fig:intro_load_net}. We have also since taken room temperature loss measurements of HDPE and UHMWPE and found that they have similar loss tangents. As such, the window loss tangent in these models is assumed to be 1.4$\pow{-4}$ unless otherwise specified \citep{Elwood2024}. In Figure \ref{fig:B3_net_loss_diff} we show how varied window loss tangents around 1.4$\pow{-4}$ affect the final calculated NET and how this compares to the measured median yearly NETs. Different loss or scattering properties between HDPE and UHMWPE may explain the increase in noise seen from 2018 to 2019, when the window material was changed.

A reduction in the white noise of a receiver directly corresponds to a reduction of integration time to reach a target sensitivity. The integration time of the receiver scales as  $t_{\text{int}} \propto \text{NET}^2$; an improvement of noise in BICEP3 of 6\% (from 267 $\mu$K$\sqrt{\text{s}}$ in 2019 thru 2021 to 252 $\mu$K$\sqrt{\text{s}}$ in 2023 and 2024) allows the receiver to map approximately 13\% faster. At higher frequencies, because of the greater fractional load reduction from thin windows, we expect this improvement in mapping speed to be even larger. At a band center of 150 GHz, a thin 1.4 mm window should decrease the NET 12\% compared to a 24.6 mm slab with the same loss tangent (from  319 $\mu$K$\sqrt{\text{s}}$ to 281 $\mu$K$\sqrt{\text{s}}$), resulting in a 22\% increase in mapping speed. We are currently analyzing the noise data from BA150 (deployed with a thin window in 2022) to confirm this expected improvement. The highest frequency bands will have an even greater improvement: at band centers 220 and 270 GHz we expect a 14\% change in NET (575 $\mu$K$\sqrt{\text{s}}$ from 668 $\mu$K$\sqrt{\text{s}}$ and 1073 $\mu$K$\sqrt{\text{s}}$ from 1251 $\mu$K$\sqrt{\text{s}}$, respectively), and 26\% faster mapping speed. When BA220/270 deploys and begins observing in the coming austral summer, we also anticipate this reduction in noise.

\section{Conclusion}

Laminated high modulus polyethylene windows provide a unique opportunity to reduce instrumental loading on millimeter wavelength telescopes. The reflection loss of the thin window can be reduced over a broad bandwidth by tuning the thickness of the window itself, providing a valuable opportunity to reduce the net reflectivity of the window to less than 0.1\% with single layer anti-reflection coats. Birefringence through the HMPE window is shown to be negligible.

We have performed extensive mechanical tests on HMPE laminates. The mechanical safety factors of the thin windows are at least 5.7 at sea level, and 8.2 at operating altitude. Though the windows do deflect significantly, they are not expected to creep into contact with the filters below them within the next five years of continuous use. 

Thin laminate windows have been successfully deployed on two BICEP/\emph{Keck} receivers at the South Pole for two observing seasons. The reduction in white noise has been conclusively measured on BICEP3, while the analysis for the newer BA150 remains ongoing. The BICEP3 window has not caused significant changes to the polarization response of BICEP3, nor has either window adversely impacted the gas accumulation in the cryostat.

We will deploy thin windows on the next two CMB receivers sent to the South Pole; the window on the high frequency BA receiver (220/270 GHz) was deployed in 2024 and the mid-frequency BA receiver (95/155 GHz) in a subsequent season \citep{Nakato2024,Petroff2024}. The primary challenge remaining for these HMPE windows are their anti-reflection coats, partially due to the large fractional bandwidths of each receiver (0.4 for BA 220/270, 0.7 for BA 95/150) \citep{Eiben2024}. A similarly difficult AR coated HMPE window will be designed for prototype wSMA receivers, covering 190 to 380 GHz, though it will require a much smaller diameter window than modern CMB small aperture telescopes \citep{Grimes2024}.

In the future, thin HMPE windows could provide similar performance improvements for an even wider range of millimeter or far IR instruments. The high safety factors may also allow thin windows to be used for even larger apertures, though the large deflection of the laminate should be taken into account in the optomechanical design.

\section{Acknowledgments}
The BICEP/\emph{Keck} projects have been made possible through a series of grants from the National Science Foundation most recently including 2220444-2220448, 2216223, 1836010, and 1726917.
The development of antenna-coupled detector technology was supported by the JPL Research and Technology Development Fund and by NASA Grants 06-ARPA206-0040, 10-SAT10-0017, 12-SAT12-0031, 14-SAT14-0009, \& 16-SAT-16-0002. 
The development and testing of focal planes was supported by the Gordon and Betty Moore Foundation at Caltech. 
Readout electronics were supported by a Canada Foundation for Innovation grant to UBC. 
Support for quasi-optical filtering was provided by UK STFC grant ST/N000706/1. 
The computations in this paper were run on the Odyssey/Cannon cluster supported by the FAS Science Division Research Computing Group at Harvard University. 

We thank the staff of the U.S. Antarctic Program and the South Pole Station who have enabled this research. 
We thank the BICEP3 winter-overs Sam Harrison (2015), Hans Boenish (2016, 2018), Grantland Hall (2017), Ta-Lee Shue (2019), Paula Crock (2020), Calvin Tsai (2021), Karsten Look (2022), Manwei Chan (2023), and Danielle Simmons (2024) as well as the BICEP Array winter-overs Thomas Leps (2022), Anthony DeCicco (2023) and Thibault Romand (2024) for their invaluable work maintaining our telescopes throughout the winter observing season.
We thank Marion Dierickx for her significant contributions during her tenure as the BICEP Operations Manager.

We also thank Vedant Chandra, Victoria DiTomasso, Christina Eiben, Dagmar Eiben, Amanda MacLean, Kevin Ortiz Ceballos, Locke Patton, Andrew Saydjari, and Rebecca Woody for their generous feedback over the years of work described in this paper.


\bibliography{text}

\end{document}

%% file: authors.tex
\author{BICEP/\textit{Keck} Collaboration:~P.~A.~R.~Ade}
\affiliation{School of Physics and Astronomy, Cardiff University, Cardiff, CF24 3AA, United Kingdom}

\author[0000-0002-9957-448X]{Z.~Ahmed}
\affiliation{Kavli Institute for Particle Astrophysics and Cosmology, Stanford University, Stanford, CA 94305, USA}
\affiliation{SLAC National Accelerator Laboratory, Menlo Park, CA 94025, USA}

\author[0000-0001-6523-9029]{M.~Amiri}
\affiliation{Department of Physics and Astronomy, University of British Columbia, Vancouver, British Columbia, V6T 1Z1, Canada}

\author[0000-0002-8971-1954]{D.~Barkats}
\affiliation{Center for Astrophysics, Harvard \& Smithsonian, Cambridge, MA 02138, USA}

\author[0000-0002-3351-3078]{R.~Basu Thakur}
\affiliation{Department of Physics, California Institute of Technology, Pasadena, CA 91125, USA}

\author[0000-0001-9185-6514]{C.~A.~Bischoff}
\affiliation{Department of Physics, University of Cincinnati, Cincinnati, OH 45221, USA}

\author[0000-0003-0848-2756]{D.~Beck}
\affiliation{Department of Physics, Stanford University, Stanford, CA 94305, USA}

\author{J.~J.~Bock}
\affiliation{Department of Physics, California Institute of Technology, Pasadena, CA 91125, USA}
\affiliation{Jet Propulsion Laboratory, California Institute of Technology, Pasadena, CA 91109, USA}

\author{H.~Boenish}
\affiliation{Center for Astrophysics, Harvard \& Smithsonian, Cambridge, MA 02138, USA}

\author{V.~Buza}
\affiliation{Kavli Institute for Cosmological Physics, University of Chicago, Chicago, IL 60637, USA}

\author{K.~Carter}
\affiliation{Center for Astrophysics, Harvard \& Smithsonian, Cambridge, MA 02138, USA}

\author[0000-0002-1630-7854]{J.~R.~Cheshire IV}
\affiliation{Minnesota Institute for Astrophysics, University of Minnesota, Minneapolis, MN 55455, USA}
\affiliation{Department of Physics, California Institute of Technology, Pasadena, CA 91125, USA}

\author{J.~Connors}
\affiliation{National Institute of Standards and Technology, Boulder, CO 80305, USA}

\author[0000-0002-2088-7345]{J.~Cornelison}
\affiliation{Center for Astrophysics, Harvard \& Smithsonian, Cambridge, MA 02138, USA}

\author{L.~Corrigan}
\affiliation{Center for Astrophysics, Harvard \& Smithsonian, Cambridge, MA 02138, USA}

\author{M.~Crumrine}
\affiliation{School of Physics and Astronomy, University of Minnesota, Minneapolis, MN 55455, USA}

\author{S.~Crystian}
\affiliation{Center for Astrophysics, Harvard \& Smithsonian, Cambridge, MA 02138, USA}

\author{A.~J.~Cukierman}
\affiliation{Department of Physics, Stanford University, Stanford, CA 94305, USA}

\author{E.~Denison}
\affiliation{National Institute of Standards and Technology, Boulder, CO 80305, USA}


\author{L.~Duband}
\affiliation{Service des Basses Temperatures, Commissariat a l’Energie Atomique, 38054 Grenoble, France}

\author{M.~Echter}
\affiliation{Center for Astrophysics, Harvard \& Smithsonian, Cambridge, MA 02138, USA}

\author[0009-0007-6718-1730]{M.~Eiben}
\correspondingauthor{Miranda~Eiben}
\email{mirandaeiben@gmail.com}
\affiliation{Center for Astrophysics, Harvard \& Smithsonian, Cambridge, MA 02138, USA}

\author[0000-0003-4117-6822]{B.~D.~Elwood}
\affiliation{Department of Physics, Harvard University, Cambridge, MA 02138, USA}
\affiliation{Center for Astrophysics, Harvard \& Smithsonian, Cambridge, MA 02138, USA}

\author[0000-0002-3790-7314]{S.~Fatigoni}
\affiliation{Department of Physics, California Institute of Technology, Pasadena, CA 91125, USA}

\author[0000-0001-8217-6832]{J.~P.~Filippini}
\affiliation{Department of Physics, University of Illinois at Urbana-Champaign, Urbana, IL 61801, USA}
\affiliation{Department of Astronomy, University of Illinois at Urbana-Champaign, Urbana, IL 61801, USA}

\author{A.~Fortes}
\affiliation{Department of Physics, Stanford University, Stanford, CA 94305, USA}

\author{M.~Gao}
\affiliation{Department of Physics, California Institute of Technology, Pasadena, CA 91125, USA}

\author{C.~Giannakopoulos}
\affiliation{Department of Physics, University of Cincinnati, Cincinnati, OH 45221, USA}

\author{N.~Goeckner-Wald}
\affiliation{Department of Physics, Stanford University, Stanford, CA 94305, USA}

\author[0000-0001-5268-8423]{D.~C.~Goldfinger}
\affiliation{Department of Physics, Stanford University, Stanford, CA 94305, USA}

\author{J.~A.~Grayson}
\affiliation{Department of Physics, Stanford University, Stanford, CA 94305, USA}

\author{A.~Greathouse}
\affiliation{Department of Physics, Stanford University, Stanford, CA 94305, USA}

\author[0000-0001-9292-6297]{P.~K.~Grimes}
\affiliation{Center for Astrophysics, Harvard \& Smithsonian, Cambridge, MA 02138, USA}

\author{G.~Hall}
\affiliation{School of Physics and Astronomy, University of Minnesota, Minneapolis, MN 55455, USA}
\affiliation{Department of Physics, Stanford University, Stanford, CA 94305, USA}

\author[0000-0003-2221-3018]{G.~Halal}
\affiliation{Department of Physics, Stanford University, Stanford, CA 94305, USA}

\author{M.~Halpern}
\affiliation{Department of Physics and Astronomy, University of British Columbia, Vancouver, British Columbia, V6T 1Z1, Canada}

\author{E.~Hand}
\affiliation{Department of Physics, University of Cincinnati, Cincinnati, OH 45221, USA}

\author{S.~A.~Harrison}
\affiliation{Center for Astrophysics, Harvard \& Smithsonian, Cambridge, MA 02138, USA}

\author{S.~Henderson}
\affiliation{Kavli Institute for Particle Astrophysics and Cosmology, Stanford University, Stanford, CA 94305, USA}
\affiliation{SLAC National Accelerator Laboratory, Menlo Park, CA 94025, USA}

\author{J.~Hubmayr}
\affiliation{National Institute of Standards and Technology, Boulder, CO 80305, USA}

\author[0000-0001-5812-1903]{H.~Hui}
\affiliation{Department of Physics, California Institute of Technology, Pasadena, CA 91125, USA}

\author{K.~D.~Irwin}
\affiliation{Department of Physics, Stanford University, Stanford, CA 94305, USA}

\author[0000-0002-3470-2954]{J.~H.~Kang}
\affiliation{Department of Physics, California Institute of Technology, Pasadena, CA 91125, USA}

\author[0000-0002-5215-6993]{K.~S.~Karkare}
\affiliation{Kavli Institute for Particle Astrophysics and Cosmology, Stanford University, Stanford, CA 94305, USA}
\affiliation{SLAC National Accelerator Laboratory, Menlo Park, CA 94025, USA}

\author{S.~Kefeli}
\affiliation{Department of Physics, California Institute of Technology, Pasadena, CA 91125, USA}

\author[0009-0003-5432-7180]{J.~M.~Kovac}
\affiliation{Department of Physics, Harvard University, Cambridge, MA 02138, USA}
\affiliation{Center for Astrophysics, Harvard \& Smithsonian, Cambridge, MA 02138, USA}

\author{C.~Kuo}
\affiliation{Department of Physics, Stanford University, Stanford, CA 94305, USA}

\author[0000-0002-6445-2407]{K.~Lau}
\affiliation{Department of Physics, California Institute of Technology, Pasadena, CA 91125, USA}

\author{M.~Lautzenhiser}
\affiliation{Department of Physics, University of Cincinnati, Cincinnati, OH 45221, USA}

\author{A.~Lennox}
\affiliation{Department of Astronomy, University of Illinois at Urbana-Champaign, Urbana, IL 61801, USA}

\author[0000-0001-5677-5188]{T.~Liu}
\affiliation{Department of Physics, Stanford University, Stanford, CA 94305, USA}

\author{K.~G.~Megerian}
\affiliation{Jet Propulsion Laboratory, California Institute of Technology, Pasadena, CA 91109, USA}

\author{M.~Miller}
\affiliation{Center for Astrophysics, Harvard \& Smithsonian, Cambridge, MA 02138, USA}

\author{L.~Minutolo}
\affiliation{Department of Physics, California Institute of Technology, Pasadena, CA 91125, USA}

\author[0000-0002-4242-3015]{L.~Moncelsi}
\affiliation{Department of Physics, California Institute of Technology, Pasadena, CA 91125, USA}

\author{Y.~Nakato}
\affiliation{Department of Physics, Stanford University, Stanford, CA 94305, USA}

\author{H.~T.~Nguyen}
\affiliation{Jet Propulsion Laboratory, California Institute of Technology, Pasadena, CA 91109, USA}
\affiliation{Department of Physics, California Institute of Technology, Pasadena, CA 91125, USA}

\author{R.~O'brient}
\affiliation{Jet Propulsion Laboratory, California Institute of Technology, Pasadena, CA 91109, USA}
\affiliation{Department of Physics, California Institute of Technology, Pasadena, CA 91125, USA}

\author{S.~Paine}
\affiliation{Center for Astrophysics, Harvard \& Smithsonian, Cambridge, MA 02138, USA}

\author{A.~Patel}
\affiliation{Department of Physics, California Institute of Technology, Pasadena, CA 91125, USA}

\author[0000-0002-4436-4215]{M.~A.~Petroff}
\affiliation{Center for Astrophysics, Harvard \& Smithsonian, Cambridge, MA 02138, USA}

\author[0000-0002-7822-6179]{A.~R.~Polish}
\affiliation{Department of Physics, Harvard University, Cambridge, MA 02138, USA}
\affiliation{Center for Astrophysics, Harvard \& Smithsonian, Cambridge, MA 02138, USA}

\author{T.~Prouve}
\affiliation{Service des Basses Temperatures, Commissariat a l’Energie Atomique, 38054 Grenoble, France}

\author[0000-0003-3983-6668]{C.~Pryke}
\affiliation{School of Physics and Astronomy, University of Minnesota, Minneapolis, MN 55455, USA}

\author{C.~D.~Reintsema}
\affiliation{National Institute of Standards and Technology, Boulder, CO 80305, USA}

\author{T.~Romand}
\affiliation{Department of Physics, California Institute of Technology, Pasadena, CA 91125, USA}

\author{D.~Santalucia}
\affiliation{Center for Astrophysics, Harvard \& Smithsonian, Cambridge, MA 02138, USA}

\author{A.~Schillaci}
\affiliation{Department of Physics, California Institute of Technology, Pasadena, CA 91125, USA}

\author{B.~Schmitt}
\affiliation{Center for Astrophysics, Harvard \& Smithsonian, Cambridge, MA 02138, USA}

\author{E.~Sheffield}
\affiliation{Center for Astrophysics, Harvard \& Smithsonian, Cambridge, MA 02138, USA}

\author[0000-0001-7387-0881]{B.~Singari}
\affiliation{School of Physics and Astronomy, University of Minnesota, Minneapolis, MN 55455, USA}

\author{K. Sjoberg}
\affiliation{Center for Astrophysics, Harvard \& Smithsonian, Cambridge, MA 02138, USA}

\author{A.~Soliman}
\affiliation{Jet Propulsion Laboratory, California Institute of Technology, Pasadena, CA 91109, USA}
\affiliation{Department of Physics, California Institute of Technology, Pasadena, CA 91125, USA}

\author{T.~St Germaine}
\affiliation{Center for Astrophysics, Harvard \& Smithsonian, Cambridge, MA 02138, USA}

\author[0000-0003-0260-605X]{A.~Steiger}
\affiliation{Department of Physics, California Institute of Technology, Pasadena, CA 91125, USA}

\author{B.~Steinbach}
\affiliation{Department of Physics, California Institute of Technology, Pasadena, CA 91125, USA}

\author{R.~Sudiwala}
\affiliation{School of Physics and Astronomy, Cardiff University, Cardiff, CF24 3AA, United Kingdom}

\author{K.~L.~Thompson}
\affiliation{Department of Physics, Stanford University, Stanford, CA 94305, USA}
\affiliation{Kavli Institute for Particle Astrophysics and Cosmology, Stanford University, Stanford, CA 94305, USA}

\author{C.~Tsai}
\affiliation{Center for Astrophysics, Harvard \& Smithsonian, Cambridge, MA 02138, USA}

\author[0000-0002-1851-3918]{C.~Tucker}
\affiliation{School of Physics and Astronomy, Cardiff University, Cardiff, CF24 3AA, United Kingdom}

\author{A.~D.~Turner}
\affiliation{Jet Propulsion Laboratory, California Institute of Technology, Pasadena, CA 91109, USA}

\author[0000-0002-3942-1609]{C.~Vergès}
\affiliation{Center for Astrophysics, Harvard \& Smithsonian, Cambridge, MA 02138, USA}

\author{A.~G.~Vieregg}
\affiliation{Kavli Institute for Cosmological Physics, University of Chicago, Chicago, IL 60637, USA}

\author[0000-0002-8232-7343]{A.~Wandui}
\affiliation{Department of Physics, California Institute of Technology, Pasadena, CA 91125, USA}

\author{A.~C.~Weber}
\affiliation{Jet Propulsion Laboratory, California Institute of Technology, Pasadena, CA 91109, USA}

\author[0000-0002-6452-4693]{J.~Willmert}
\affiliation{School of Physics and Astronomy, University of Minnesota, Minneapolis, MN 55455, USA}

\author[0000-0001-5411-6920]{W.~L.~K.~Wu}
\affiliation{SLAC National Accelerator Laboratory, Menlo Park, CA 94025, USA}
\affiliation{Kavli Institute for Particle Astrophysics and Cosmology, Stanford University, Stanford, CA 94305, USA}

\author{H.~Yang}
\affiliation{Department of Physics, Stanford University, Stanford, CA 94305, USA}

\author[0000-0002-8542-232X]{C.~Yu}
\affiliation{Kavli Institute for Cosmological Physics, University of Chicago, Chicago, IL 60637, USA}

\author[0000-0001-6924-9072]{L.~Zeng}
\affiliation{Center for Astrophysics, Harvard \& Smithsonian, Cambridge, MA 02138, USA}

\author[0000-0001-8288-5823]{C.~Zhang}
\affiliation{Department of Physics, California Institute of Technology, Pasadena, CA 91125, USA}

\author{S.~Zhang}
\affiliation{Department of Physics, California Institute of Technology, Pasadena, CA 91125, USA}
